\documentclass[reprint,aps,prb,amsmath,amssymb,longbibliography,superscriptaddress]{revtex4-2}
\usepackage{bm}
\usepackage{graphicx}
\DeclareUnicodeCharacter{2212}{\textminus}
\usepackage[colorlinks, linkcolor=blue, citecolor=blue, urlcolor=blue]{hyperref}

\begin{document}
	\title{Quantum-metric Bloch oscillations in weakly inhomogeneous electric fields}
	
	\author{M. Maneesh Kumar}
	\email{mmkr20@iitk.ac.in}
	\affiliation{Department of Physics, Indian Institute of Technology Kanpur, Kanpur 208016}
	\author{Md Kaif Faiyaz}
	\affiliation{Department of Physics, Indian Institute of Technology Kanpur, Kanpur 208016}
	\author{Sayan Sarkar}
	\affiliation{Department of Physics, Indian Institute of Technology Kanpur, Kanpur 208016}
	\author{Amit Agarwal}
	\email{amitag@iitk.ac.in}
	\affiliation{Department of Physics, Indian Institute of Technology Kanpur, Kanpur 208016}
	
	\begin{abstract}
		Geometric analogs of Bloch oscillations studied so far have relied on Berry curvature. We show that a weakly inhomogeneous electric field adds a distinct quantum-metric term to semiclassical wavepacket dynamics, generating an oscillatory real-space contribution even when the Berry curvature vanishes. The associated transport response comprises an intrinsic and a scattering-time-dependent part. In the regime studied, the latter can dominate and approach finite saturation at high field when the relative field inhomogeneity is held fixed. A tilted Dirac model illustrates the mechanism. Realistic platforms will likely require synthetically engineered superlattices, with a finite quantum metric and an adequate band gap.
	\end{abstract}
	\maketitle
	\section{Introduction}
	
	Bloch oscillations are the coherent periodic motion of an electron wavepacket in a periodic potential under a constant force field. Predicted by Felix Bloch in 1929 \cite{bloch1929}, they remain a canonical example of wave-packet dynamics in crystals. In momentum space, the electron traverses a band under a weak static field. In real space, that motion appears as an oscillation rather than sustained translation across the lattice \cite{Longhi2009, Price2012}. Their experimental observation in solids is difficult because dephasing, narrow tunneling windows, and the small lattice periods of natural crystals severely constrain the accessible oscillation scales \cite{Gustavsson2008, Unuma2006, Meinert2014, Fahimniya2021}. Even so, Bloch oscillations and related effects have been studied extensively \cite{Dekorsy1995, Krieger1986, tsu1973, averin1985}, and electronic Bloch oscillations or their analogs have been explored in systems ranging from topological insulators \cite{DiLiberto2020} and strongly correlated quantum liquids \cite{Meinert2017} to optical lattices \cite{Preiss2015, Qin2024}, cold atoms \cite{Pagel2020}, photonic lattices \cite{Lebugle2015, Corrielli2013, Xu2016}, gases \cite{Floss2015}, micro-resonators \cite{Chen2021}, and 1D chains \cite{Merlin2024}.
	
	Recent work proposed a geometric analog of Bloch oscillations \cite{Phong2023, Beule2023}. In that setting, the oscillatory motion comes from the anomalous velocity and is therefore controlled by the Berry curvature of the Bloch band. This provides a way to probe quantum geometry through Bloch-like dynamics and may also be relevant for accessing the terahertz regime \cite{Davies2004, Borak2005}. But Berry-curvature-driven oscillations are symmetry-restricted. In systems preserving combined parity-time-reversal symmetry ($\mathcal{PT}$), the Berry curvature vanishes pointwise throughout the Brillouin zone, and the corresponding geometric oscillation is absent.
	\begin{figure}
		\centering
		\includegraphics[width=0.9\linewidth]{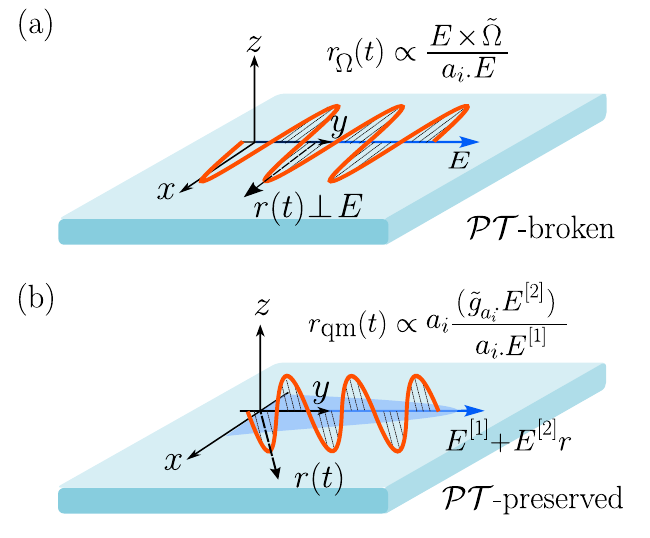}
		\caption{(a) Schematic of Berry-curvature-driven Bloch oscillations in $\mathcal{PT}$-broken systems. The wavepacket oscillates perpendicular to the applied homogeneous field ($E$). (b) Quantum-metric-induced Bloch oscillations in the presence of an inhomogeneous electric field. The resulting motion can have components both parallel and perpendicular to the applied field.}
		\label{fig_schematic}
	\end{figure}
	This raises a natural question: can a band-geometric analog of Bloch oscillations survive even when the Berry curvature vanishes? One possibility is to use nonlinear regimes to achieve this. Here, instead, we demonstrate this possibility using weakly inhomogeneous \cite{ZengPRB2025} driving fields. In this case, the semiclassical equations of motion acquire explicit quantum-metric contributions in addition to the Berry-curvature terms \cite{Lapa2019}, making it possible to isolate a quantum-metric-driven part of the wavepacket dynamics in systems with vanishing Berry curvature.
	
	Here we develop a semiclassical theory of Bloch-electron dynamics in a weakly inhomogeneous electric field and isolate the quantum-metric contribution to the oscillatory motion. The result is a Bloch-like oscillatory term that survives even when the Berry curvature vanishes. Our aim is not to rederive the inhomogeneous-field semiclassical framework itself, but to extract the oscillatory and transport consequences of its quantum-metric term. The same mechanism generates intrinsic and extrinsic transport contributions, and in the regime studied the extrinsic current can dominate and approach finite high-field saturation. We illustrate the effect with a tilted Dirac model that has vanishing Berry curvature and finite quantum metric. Realistic experimental platforms, however, will require larger-unit-cell systems with an adequate band gap.
	
	The paper is structured as follows. In Sec.~\ref{section_two}, we introduce the inhomogeneous-field semiclassical framework and derive the quantum-metric contribution to the wavepacket dynamics. In Sec.~\ref{section_three}, we turn to the associated transport response and separate its intrinsic and extrinsic pieces, with emphasis on the strong-field behavior. In Sec.~\ref{section_four}, we use a tilted Dirac model to illustrate the symmetry conditions and scaling of the response. Finally, in Sec.~\ref{Section_five}, we summarize the main results and their scope.
	
	\section{Quantum-metric contribution to real-space oscillations}\label{section_two}
	
	The semiclassical equations of motion provide the natural starting point for identifying how weak spatial inhomogeneity modifies Bloch-electron dynamics \cite{ashcroft1976}. Our goal in this section is to isolate the quantum-metric contribution generated by a weakly inhomogeneous electric field and to determine the oscillatory component of the corresponding real-space motion. To include such inhomogeneous perturbations, we work with the extended semiclassical framework appropriate to slowly varying external fields. The Hamiltonian of the perturbed system, taken in the length gauge, assumes the form \cite{Aversa1995, Kumar2024}, 
	\begin{equation}    
		\hat{\mathcal{H}} = \hat{\mathcal{H}}_0 - e \hat{\phi}(\hat{{\bm{r}}}).
	\end{equation}
	Here, the first term $\hat{\mathcal{H}}_0 = (1/2m)\delta^{a b}\hat{p}_a\hat{p}_b + \hat{V}(\hat{{\bm{r}}}),$ represents a Bloch electron of mass $m$ and charge $-e$ in a periodic potential, $\hat{V}(\hat{{\bm{r}}})$. In the weak-field regime, the second term is a perturbative external potential of the form $\hat{\phi}(\hat{{\bm{r}}}) = -E_a \hat{r}^a - (1/2) E_{a b}\hat{r}^a \hat{r}^b$.
	We define $\hat{r}^a$ and $\hat{p}_a$ as the single-electron position and momentum operators, respectively, with $[\hat{r}^a,\hat{p}_b] = i\hbar \delta^a_b$, where $\delta^a_b$ (or $\delta^{ab}$) is the Kronecker delta. The electric field perturbation, $\tilde{E}_a=-\partial_{r^a}\phi({\bm{r}}) = E_a + E_{a b}r^b$, consists of a static field term, $E_a$, and a finite field-gradient term, $E_{a b} r^b$. Here $E_a$ and $E_{ab}$ are constant parameters with $E_{ab} = E_{ba}$. The modified semiclassical equations of motion, including the inhomogeneous-field corrections, are given by \cite{Lapa2019, kumar2025},
	\begin{align}
		\Dot{r}^a &= \frac{1}{\hbar}\partial_{k_a} \epsilon_{\bm{k}} - \Omega^{a b}_{\bm{k}}\Dot{k}_b + \frac{e}{2\hbar}E_{b d}\partial_{k_a} g^{b d}_{\bm{k}},\label{x_dot}\\
		{\hbar}\Dot{k}_b &= -e (E_{b} + E_{b c}r^c).\label{k_dot}
	\end{align}
	Here, $\epsilon_{\bm{k}}$ is the unperturbed band dispersion, and $\Omega^{ab}_{\bm{k}}$ and $g^{ab}_{\bm{k}}$ are the single-band Berry curvature and quantum metric matrix elements, with $a,b,c,d ~ \in \{x,y\}$ for a 2D system. The quantum metric and Berry curvature are, respectively, the real and imaginary parts of the quantum geometric tensor. They encode the distance between neighboring Bloch states in momentum space and the self-rotation of the wavepacket as it traverses parameter space.
	From Eqs. \eqref{x_dot} and \eqref{k_dot}, the decoupled expression for the band velocity takes the form,
	\begin{equation}\label{net_vel}
		\Dot{r}^a = \frac{1}{\hbar} \partial_{k_a}\epsilon_{\bm{k}} + \frac{e}{\hbar} \Omega^{a b}_{\bm{k}}(E_{b} + E_{b c}r^c) + \frac{e}{2\hbar} E_{bd}\partial_{k_a}g^{bd}_{\bm{k}}. 
	\end{equation}
	We discuss the details of the derivation in Appendix~\ref{app_sem_cl}. In Eq.~\eqref{net_vel}, the first term gives rise to the conventional Bloch oscillation \cite{Bonilla2011}. The Berry-curvature term contains both the usual anomalous velocity responsible for the geometric oscillations of Ref.~[\onlinecite{Phong2023}] and a gradient-coupled contribution arising from $E_{bc}r^c$. The last term is the quantum-metric contribution and gives rise to what we term the quantum-metric oscillation. The physics surrounding this contribution is the focus of our paper. Among these responses, the Bloch oscillation term is band-dispersive in origin, while the Berry-curvature and quantum-metric terms originate from the quantum geometry of the Bloch electron.
	
	In this work, we restrict our attention to systems with vanishing Berry curvature so that the quantum-metric term in Eq.~\eqref{net_vel} can be isolated from anomalous-velocity effects. Symmetry-wise, this includes systems with $\mathcal{PT}$ symmetry, for which the Berry curvature vanishes identically throughout the Brillouin zone. To obtain the real-space motion, we solve Eq.~\eqref{net_vel} by Fourier expanding the band dispersion and quantum metric as ${\epsilon}_{\bm{k}} = \sum_{\bm{a}_i} e^{i\bm{k}\cdot\bm{a}_i}\tilde{{\epsilon}}_{\bm{a}_i}$ and $\bm{g}_{\bm{k}} = \sum_{\bm{a}_i} e^{i\bm{k}\cdot\bm{a}_i}\tilde{\bm{g}}_{\bm{a}_i}$, where $\bm{a}_i$ are lattice vectors. Using the corresponding time dependence of $\bm{k}(t)$ from Eq.~\eqref{k_dot}, the resulting $\bm{r}(t)$ separates naturally into a drift part and an oscillatory part. The drift term arises when $\bm{a}_i\cdot \tilde{\bm{E}}= E_a a^{a}_i +  E_{a b}r^b a^{a}_i=0$ (see Appendix~\ref{osc_calc}) and contains both band-dispersive and band-geometric contributions. The oscillatory part, by contrast, is obtained from the nonorthogonal terms and is where the quantum-metric-driven dynamics appears most directly.
	The oscillatory part of the wavepacket position consists of the conventional Bloch term, ${\bm r}_{\text{Bloch}}$, and the quantum-metric term, $ \bm{r}_{\text{qm}}$, given by
	\begin{widetext}
		\begin{subequations}
			\begin{eqnarray}   
				\bm{r}_{\text{Bloch}}(t) &=&  \sum_{\tilde{\bm{E}}\cdot \bm{a}_i \neq 0}  e^{i\bm{k}_{0}\cdot \bm{a}_i}[1 - \exp{(-ie \bm{E}^{[1]}\cdot \bm{a}_i t/\hbar)}]\frac{\bm{a}_i \,  \tilde{{\epsilon}}_{\bm{a}_i}}{e\bm{E}^{[1]} \cdot \bm{a}_i} , \label{bloch_osc_mt}\\
				\bm{r}_{\text{qm}}(t) &=&  \frac{1}{2}\sum_{\tilde{\bm{E}}\cdot \bm{a}_i \neq 0} e^{i\bm{k}_{0}\cdot \bm{a}_i} [1 - \exp{(-ie \bm{E}^{[1]} \cdot \bm{a}_it/\hbar)} ] \frac{ \bm{a}_i \, ({\tilde{\bm{g}}_{\bm{a}_i}}\cdot \bm{E}^{[2]})}{ \bm{E}^{[1]} \cdot \bm{a}_i}.\label{qme_osc_mt}
			\end{eqnarray}    
		\end{subequations}
	\end{widetext}
	
	Here, $\bm{k}_0=\bm{k}(t=0)$ denotes the initial wave vector, while $\bm{E}^{[1]}$ and $\bm{E}^{[2]}$ denote the static and gradient components of the applied electric field, with matrix elements $E_i$ and $E_{ij}$, respectively. Throughout, we assume the gradient component to be weaker than the uniform one, so that $\tilde{\bm{E}}\cdot\bm{a}_i \simeq \bm{E}^{[1]}\cdot\bm{a}_i$. Eqs.~\eqref{bloch_osc_mt} and~\eqref{qme_osc_mt} then show explicitly that the field gradient generates an additional oscillatory contribution controlled by the quantum metric of the Bloch band.
	These expressions apply when the field varies slowly over the wavepacket size and over the spatial excursion of the orbit, so that $|E_{ab}|\ell_{\rm wp}\ll |E_a|$ and $|E_{ab}|A_{\rm osc}\ll |E_a|$, where $\ell_{\rm wp}$ is the wavepacket width and $A_{\rm osc}$ is the oscillation amplitude. The dynamics is single-band and adiabatic. The work done over a lattice spacing and over the wavepacket size must remain small compared with the relevant interband gap, so that Landau-Zener tunneling is negligible. In addition, the transport formulas below use a local relaxation-time approximation with a field-independent $\tau$, neglect spatial gradients of the equilibrium distribution, and treat the measured current as the sample average of the local current density. The strong-field saturation discussed later should therefore be read in the convention where the relative inhomogeneity $E_{ab}a/|E_a|$ is held fixed while the overall field scale is increased.
	Eq.~\eqref{bloch_osc_mt} shows that the amplitude of conventional Bloch oscillations grows inversely with the applied field and therefore becomes large in the weak-field regime. In that limit the electron must traverse a long real-space distance to complete one oscillation, which makes the motion especially vulnerable to interruption by impurity, phonon, or defect scattering once the oscillation amplitude exceeds the mean free path. This is one reason why Bloch oscillations remain difficult to observe in ordinary crystals. In contrast, Eq.~\eqref{qme_osc_mt} shows that the quantum-metric contribution scales with the ratio of the field gradient to the static field. This keeps the oscillation amplitude bounded by the relative field inhomogeneity and correspondingly reduces the spatial excursion of the orbit. The resulting motion is therefore less exposed to disorder-induced interruption and is, in principle, more accessible experimentally.
	
	To summarize, the inhomogeneous-field semiclassical dynamics contains, in addition to the conventional Bloch term, a quantum-metric contribution that generates an oscillatory component even when the Berry curvature vanishes. Having identified the dynamical origin of this effect, we now turn to the associated current response, which provides the more accessible experimental signature.
	\begin{figure*}
		\centering    
		\includegraphics[scale = 0.3]{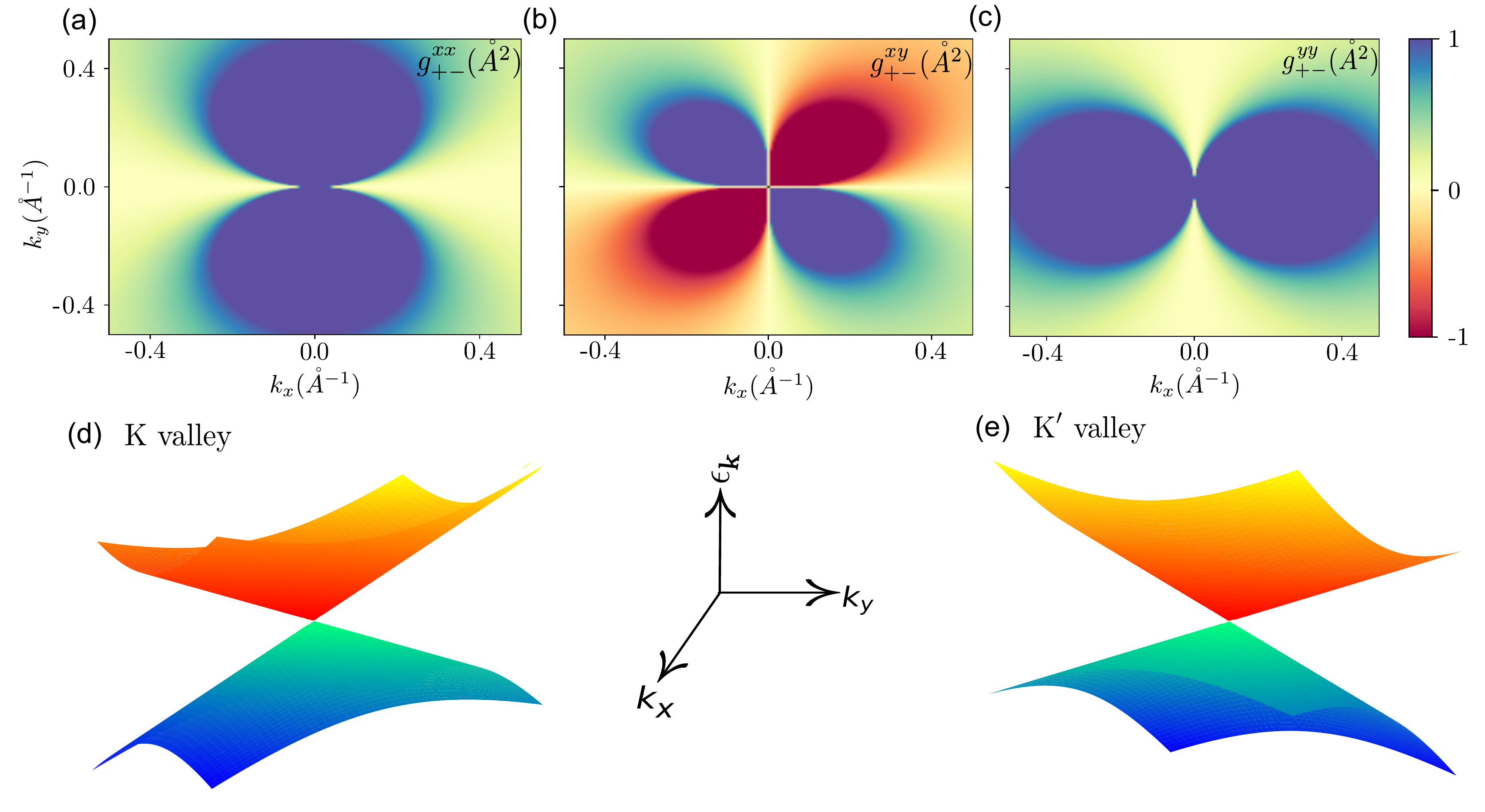}
		\caption{Panels (a)-(c) show the interband quantum-metric components $g^{xx}_{+-}$, $g^{xy}_{+-}$, and $g^{yy}_{+-}$, respectively, with $+$ ($-$) denoting the conduction (valence) band. Panels (d) and (e) show the tilted Dirac cones at the $\rm K$ and $\rm K'$ valleys, corresponding to $\eta=+1$ and $\eta=-1$ following Eq.~\eqref{Ham}. The parameters are $v_t = 0.2v_F$ and $\hbar v_F = 1$ eV.}    
		\label{paper_figure1}
	\end{figure*}
	
	\section{Transport signatures of quantum-metric oscillatory dynamics}\label{section_three}
	
	Directly tracking the wavepacket motion described above is generally more difficult than probing its transport consequences. We therefore now focus on the current response associated with the Bloch and quantum-metric contributions to the dynamics. Our main interest is the quantum-metric-driven transport channel and, in particular, the regime where it exhibits finite saturation. In the semiclassical picture, the local charge current density is defined as \cite{Bhalla2022, Varshney2023}
	\begin{equation}\label{c_density}
		\bm{j} (\bm{R}) = -e \int_{\bm{r}} \int_{\bm{k}} \, \Dot{{\bm{r}}}~f \delta(\bm{r} - \bm{R}).
	\end{equation}
	Here, $\int_{\bm{k}} = \int d\bm{k}/(2\pi)^D$, where $\bm{k} = (k_x,k_y,\dots)$ is the $D$-dimensional Bloch wavevector. We also define $\int_{\bm{r}} = \int_{\text{s}} d\bm{r}$, with $s$ spanning the sample, $\bm{R}$ denoting the observation point, and $f = f_{\bm{k}}$ the nonequilibrium distribution function (NDF).
	The NDF is obtained by solving the semiclassical Boltzmann equation (SBE),
	\begin{equation}\label{Bm_eq_coll}
		{\partial_t f } + \Dot{{\bm{r}}}\cdot \nabla_{\bm{r}}f + \Dot{\bm{k}}\cdot \nabla_{\bm{k}}f = \mathcal{I}_{\text{coll}}\{ f(\bm{r}, \bm{k})\},
	\end{equation}
	where $\mathcal{I}_{\text{coll}}\{ f(\bm{r}, \bm{k})\}$ is the collision integral \cite{Chang1995, ashcroft1976, Das2019}. A spatially varying perturbing field induces a real-space modulation of the carrier distribution, and therefore a finite $\nabla_{\bm r} f$ term. Writing $f = f_0 + \delta f$, where $\delta f$ is the correction to the equilibrium distribution function (EDF), and keeping only the leading correction in the weak-field regime, the relaxation-time approximation gives
	\begin{equation}\label{Bm_eq}
		\dot{\bm{r}} \cdot \nabla_{\bm{r}}f_0 + \dot{\bm{k}} \cdot \nabla_{\bm{k}}f_0 = -\frac{\delta f}{\tau},  
	\end{equation}
	which gives
	\begin{equation}
		\delta f \simeq -\tau(\dot{\bm{r}}\cdot\nabla_{\bm{r}}f_0 + \dot{\bm{k}}\cdot\nabla_{\bm{k}}f_0). 
	\end{equation}
	The first term captures nonequilibrium effects driven by spatial gradients, as in diffusion and heat transport \cite{mazumder2021}. The second term captures the change in occupation generated by the external force and is the relevant contribution for the electrical response considered here. We therefore work in the regime where the spatial gradient of the equilibrium distribution can be neglected, giving
	\begin{equation} \label{f1}
		\delta f \simeq -\tau (\dot{\bm{k}}\cdot \nabla_{\bm k} \epsilon_{\bm{k}}) \frac{\partial f_0}{\partial \epsilon_{\bm{k}}}.
	\end{equation}
	The scattering-time dependence identifies the extrinsic Boltzmann contribution to the distribution function. An alternative Fourier-transform solution gives a more general NDF, applicable up to larger electric-field strengths:
	\begin{equation}\label{NDF}
		f = \sum_{{\bm{a}}_i} (1- ie{\tilde{\bm{E}}}\cdot {{\bm{a}}_i \tau /\hbar})^{-1} \tilde{f}^{(0)}_{{\bm{a}}_i} e^{i{\bm{k}}\cdot{{\bm{a}}_i}}, 
	\end{equation} 
	which in the large-$\tilde{\bm{E}}$ limit reduces to
	\begin{equation}  \label{f1_largeE}
		f \simeq \sum_{\tilde{\bm{E}}\cdot\bm{a}_i =0}\tilde{f}^{(0)}_{\bm{a}_i} e^{i{\bm{k}}\cdot{\bm{a}_i}} + \frac{i\hbar}{e \tau} \sum_{\tilde{\bm{E}}\cdot\bm{a}_i\neq 0} \frac{\tilde{f}^{(0)}_{\bm{a}_i} e^{i{\bm{k}}\cdot{\bm{a}_i}}}{\tilde{\bm{E}}\cdot\bm{a}_i}, 
	\end{equation}
	where $f_0=\sum_{\bm{a}_i}\tilde{f}^{(0)}_{\bm{a}_i} e^{i{\bm{k}}\cdot{\bm{a}_i}}$. This form is useful for analyzing the differential conductance of band-dispersive and band-geometric currents in the large-field limit. For the low-field response, we use Eq.~\eqref{f1}.
	
	\subsection{Lowest order response}
	
	The steady-state drift response contains intrinsic and extrinsic mechanisms. The intrinsic contribution is determined by the band structure itself and includes both band-dispersive and band-geometric parts. The band-dispersive part corresponds to the Bloch current \cite{Phong2023},
	\begin{equation}\label{bloch_current}
		j^a_{\text{Bloch}} = -\frac{e}{\hbar}\int_{\bm{k}}  \partial_a\epsilon_{\bm k} ~f.
	\end{equation}
	Our main focus is the band-geometric current generated by the quantum metric. In the presence of spatial inhomogeneity, this current also contains intrinsic and extrinsic pieces. Within the intrinsic regime, it is given by \cite{Lapa2019}
	\begin{equation}
		j^{a}_\text{qmi} = -\frac{e^2}{2\hbar} \int_{\bm{k}} E_{d b}\partial_{k_a}g^{d b}_{\bm{k}} \, f_0. \label{j_qmi}
	\end{equation}
	Here, the real-space integral is unity. The intrinsic quantum-metric current depends only on the strength of the gradient field, unlike the corresponding extrinsic current, which also depends on the static field strength through the nonequilibrium distribution. In this work, ``extrinsic'' refers to the symmetric scattering-time-dependent Boltzmann contribution. It does not include asymmetric scattering contributions such as skew scattering or side-jump mechanisms. We therefore keep only the part of the distribution function with $\tau^1$ dependence. The $a$th component of the extrinsic quantum-metric current, ${j}^a_{\text{qme}}$, is then given by \cite{Lapa2019}
	\begin{align}
		{j}^a_{\text{qme}}(\bm{R})
		&=
		- \frac{e^2}{2\hbar}
		\int_{\bm r} \int_{\bm k}
		E_{d b}\partial_{k_a}g^{d b}_{\bm{k}}
		\delta f
		\delta(\bm{r} - \bm{R}),
		\label{j_qme}\\
		&=
		-\int_{\bm{r}}\alpha_c \tau \int_{\bm{k}}
		(E_{d b}\partial_{k_a}g^{d b}_{\bm{k}})
		(\partial_{k_c} \epsilon_{\bm{k}})
		\frac{\partial f_0}{\partial \epsilon_{\bm{k}}}
		\delta(\bm{r} - \bm{R}).
		\label{jqme_a}
	\end{align}
	with
	\begin{equation}
		\alpha_c = \frac{e^3}{2\hbar^2}(E_c + E_{c b}r^b).
	\end{equation}
	Here and below, Einstein's summation convention is used for repeated indices. The delta function in Eq.~\eqref{jqme_a} fixes the real-space coordinate in $\alpha_c$ to the observation point, so that the local response is controlled by $\alpha_c(\bm R)=e^3(E_c+E_{cb}R^b)/(2\hbar^2)$. With the convention $\int_{\bm k}=d^Dk/(2\pi)^D$, Eq.~\eqref{jqme_a} has the units of a $D$-dimensional charge-current density. The factor $E_{db}\partial_{k_a}g^{db}_{\bm k}$ has the dimensions of a velocity prefactor after multiplication by $e/\hbar$, while the extra factor $e\tau\tilde E_c\partial_{k_c}\epsilon_{\bm k}/\hbar$ is the dimensionless first-order correction to the distribution function. In the analytic expressions below, we use $h=2\pi\hbar$. For the model plots, we use lattice units in which the lattice length is set to unity, so the plotted currents should be read in the corresponding natural current-density units.
	Eq.~\eqref{j_qmi} and \eqref{jqme_a} show that the intrinsic response is a Fermi-sea effect, whereas the extrinsic response is a Fermi-surface effect \cite{Sinitsyn2005, Yao2004}. They also differ in their electric-field scaling. The intrinsic current, $\bm{j}_\text{qmi}$, is linear in the field gradient and reverses direction under $E_{ij} \rightarrow - E_{ij}$. By contrast, the extrinsic current, $\bm{j}_\text{qme}$, is nonlinear because it combines the gradient field entering the quantum-metric velocity with the field-dependent correction to the distribution function. The lowest-order extrinsic response can therefore be viewed as a second-order response involving both longitudinal and transverse components. Unlike $j_{\text{qmi}}$, it also retains an explicit dependence on real-space position through the spatially inhomogeneous perturbing field.
	
	An important aspect of Eq.~\eqref{j_qme} is its behavior in the higher-field regime, because the corresponding differential conductance provides a useful signature of the underlying dynamics. Conventional Bloch oscillations are associated with negative differential conductance \cite{Hyart2009}. This behavior originates from Wannier-Stark localization: a strong uniform electric field reorganizes a Bloch band into localized ladder states and suppresses coherent intraband transport across the lattice \cite{Esaki1970, Beltram1990, Fahimniya2021}. The quantum-metric current has a different behavior in the high-field limit. Using Eqs.~\eqref{f1_largeE} and \eqref{j_qme}, we find
	
	\begin{align}
		j^a_{\text{qme}}(\bm{R})
		&=
		\frac{e}{2i\tau}
		\int_{\bm{r}} \int_{\bm{k}}
		E_{d b}\partial_{k_a}g^{d b}_{\bm{k}}
		\sum_{\tilde{\bm{E}}\cdot\bm{a}_i\neq 0}\nonumber\\
		&\quad\times
		\frac{\tilde{f}^{(0)}_{\bm{a}_i} e^{i{\bm{k}}\cdot{\bm{a}_i}}}
		{\tilde{\bm{E}}\cdot\bm{a}_i}
		\delta(\bm{r}-\bm{R}),
	\end{align}
	In the large-field limit, $\bm{j}_{\text{qme}}$ approaches an almost field-independent regime when the relative field inhomogeneity is kept fixed. The differential conductance then vanishes rather than becoming negative. This finite geometric response can be understood from the interband Berry-connection matrix elements that define the quantum metric \cite{Phong2023}. Even when ordinary intraband dispersive transport is suppressed by Stark localization, these geometric matrix elements continue to encode the overlap structure of neighboring Bloch states. In a spatially inhomogeneous field, the ladder spectrum also acquires gradient corrections involving band-geometric quantities, so the resulting current should be viewed as a consequence of the modified geometric matrix elements and transition amplitudes rather than as a direct consequence of uniform ladder spacing alone.
	
	To summarize, this section identifies the transport signature of the quantum-metric contribution, with particular emphasis on the extrinsic response and its finite strong-field saturation. We next illustrate these general results in a simple low-energy model.
	
	\begin{figure*}
		\centering
		\includegraphics[width=0.75\linewidth]{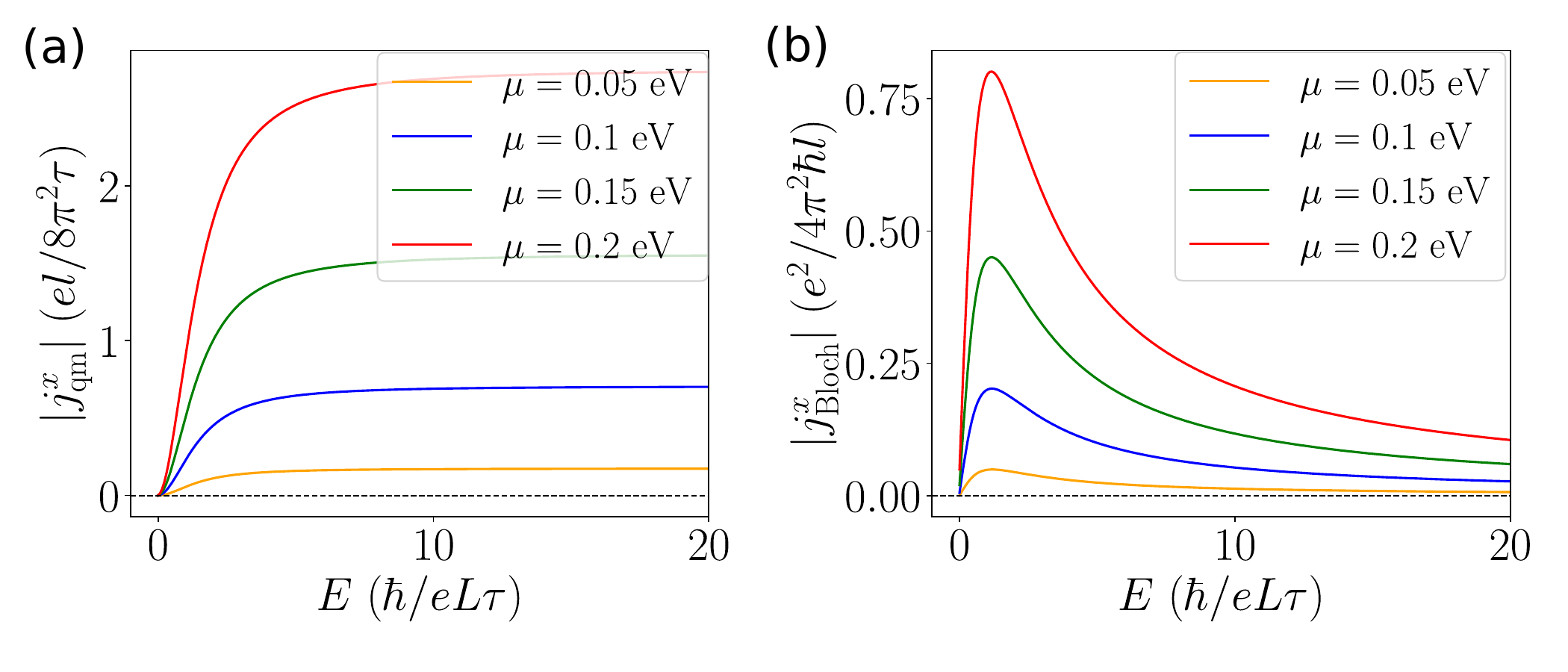}
		\caption{(a) Magnitude of the quantum-metric current $j_{\text{qm}}$ along $\hat{x}$ for the tilted Dirac model [Eq.~\eqref{Ham}] as a function of the applied field for various chemical potentials $\mu$, with an inhomogeneous electric field applied along $\hat{x}$ with a finite longitudinal gradient $E_{xx}$. The current approaches a finite saturation in the high-field limit, implying a vanishing differential conductance and providing the main transport signature discussed in the text. Here $L$ denotes the nearest-neighbor lattice spacing set by the primitive lattice vectors $\bm{a}_i$, and $l$ is the sample length along the current direction. (b) Longitudinal conventional Bloch current versus field, showing negative differential conductivity at large field, thereby signaling the underlying Wannier-Stark localization.}
		\label{qm_currentvsE}
	\end{figure*}

	\section{Illustrative low-energy model}\label{section_four}
	
	In this section, we use a simple low-energy model to illustrate the symmetry constraints and scaling of the response. Rather than making a material-specific prediction, we use a tilted Dirac model \cite{Varshney2023, Mojarro2021, Mojarro2022} to show how the quantum-metric contribution identified above appears in a concrete system with vanishing Berry curvature and finite quantum metric. Direct experimental observation of the oscillations requires a system with a sufficiently large unit cell and an adequate band gap to suppress Zener tunneling, so the present model should be viewed as illustrative rather than material specific. The low-energy Hamiltonian is given by
	\begin{equation}\label{Ham}
		\mathcal{H} = \eta v_t k_x \sigma_0 + v_F (\eta k_x\sigma_x + k_y \sigma_y), 
	\end{equation}
	where $\sigma_{x,y}$ are Pauli matrices, $v_t$ is the tilt velocity, and $v_F$ is the Fermi velocity. Figs.~\ref{paper_figure1}(d),(e) show the band structure for each valley. The Bloch eigenstates are
	\begin{equation}
		| u \rangle_s = \frac{1}{\sqrt{2}} \begin{pmatrix}
			1 \\
			s  e^{i \theta}
		\end{pmatrix},    
	\end{equation}
	where $e^{i \theta} = \frac{\eta k_x + i k_y}{\sqrt{(k^{2}_x + k^{2}_y)}}$, $\eta=\pm 1$ denotes the valley index (K/K$'$), and $s=\pm1$ denotes the conduction/valence band. The corresponding band dispersion is
	\begin{equation}
		\epsilon_{{\bm{k}}, s}
		=
		\eta v_t k_x+s\epsilon_0,
		\qquad
		\epsilon_0=v_F|{\bm{k}}|,
	\end{equation}
	which also yields the Bloch current defined in Eq.~\eqref{bloch_current}.
	The interband quantum-metric matrix elements, which give rise to the band-geometric response, are
	\begin{subequations}
		\begin{eqnarray}
			g^{xx}_{ss'} &=& \frac{v^4_Fk^2_y}{4 \epsilon^4_{0}}(1-\delta_{ss'}),\\
			g^{xy}_{ss'} &=& - \frac{k_xk_yv^4_F}{4 \epsilon^4_{0}}(1- \delta_{ss'}), \\
			g^{yy}_{ss'} &=& \frac{v^4_Fk^2_x}{4 \epsilon^4_{0}} (1-\delta_{ss'}).
		\end{eqnarray}
	\end{subequations}
	Here, $s,s' \in \{+, -\}$ denote the conduction (+) and valence ($-$) bands, respectively. Fig.~\ref{paper_figure1}(a)-(c) illustrates the momentum-space distribution of the different quantum-metric components around the $\rm K$ point. We first examine the scaling behavior of the quantum-metric current ($J_{qm}$) with the applied electric field using the full nonequilibrium distribution function in Eq.~\eqref{NDF}. For the calculations, we consider an electric field of the form $\tilde{\bm{E}} = ({E}_1+E_{11}x)~\hat{x}$ with $E_{11}L = E_1/10$, where $L$ denotes the nearest-neighbor lattice spacing, and set $v_t = 0.2v_F$, $\hbar v_F=1~\rm eV$, and $\eta=1$. The numerical evaluation uses the same tensor-contraction convention as Eq.~\eqref{Eg}. For this field geometry, the only nonzero gradient component is $E_{xx}=E_{11}$, and the relative inhomogeneity $E_{11}L/E_1$ is kept fixed while scanning the field amplitude. We use the primitive lattice vectors $\bm{a}_1 = L(-1/2,\sqrt{3}/2)$, $\bm{a}_2 = L(1,0)$, and $\bm{a}_3=-\bm{a}_1-\bm{a}_2$. As shown in Fig.~\ref{qm_currentvsE}(a), the quantum-metric current grows linearly with the electric field in the weak-field regime. As the field strength increases, the current approaches saturation, leading to a nearly vanishing differential conductance at high fields. As expected, this behavior contrasts sharply with the conventional Bloch current, which exhibits negative differential conductance, as evaluated and shown in Fig.~\ref{qm_currentvsE}(b) for the same model.

	\begin{figure*}[t]
		\centering
		\includegraphics[scale = 0.55]{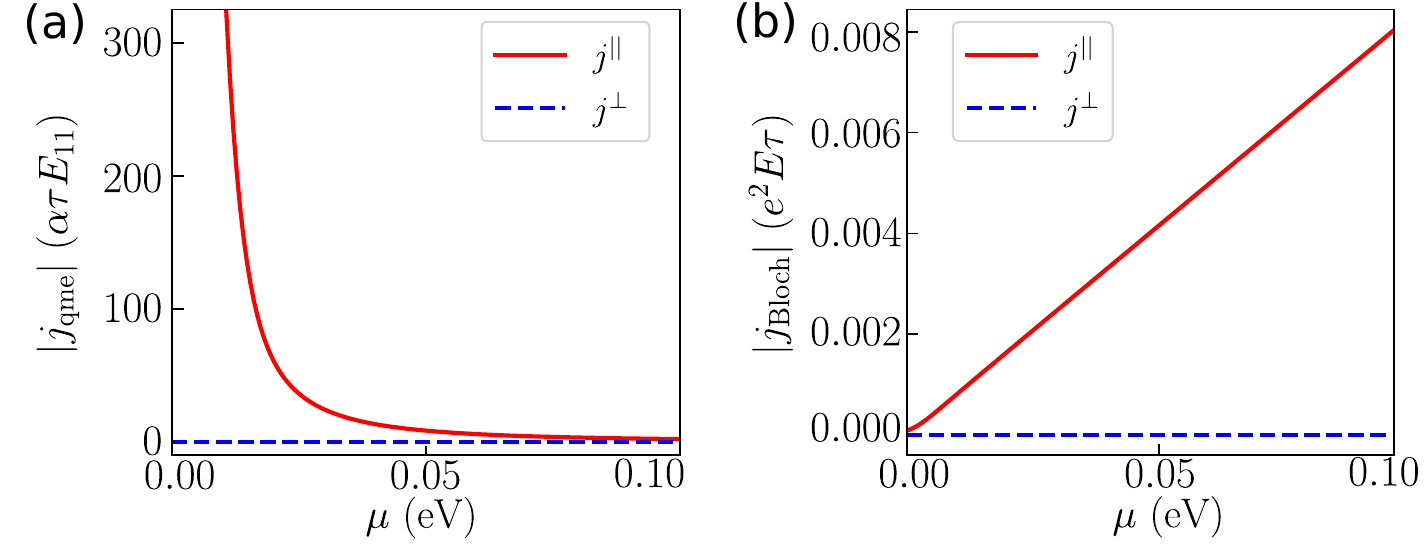}
		\caption{Panel (a) shows the longitudinal and transverse quantum-metric currents, $j^{||}_{\text{qme}}$ and $j^{\perp}_{\text{qme}}$, as functions of the chemical potential ($\mu$) for $\tilde{\bm{E}} = ({E}_1+E_{11}x)~\hat{x}$. Panel (b) shows the longitudinal and transverse Bloch currents, $j^{||}_{\text{Bloch}}$ and $j^{\perp}_{\text{Bloch}}$, as functions of the chemical potential for the same field. The figure highlights that the quantum-metric response is strongest near charge neutrality and that tilt enhances the longitudinal current. The numerical parameters are $\eta = 1$, $\hbar v_F = 1$ eV, and $v_t = 0.2 v_F$.}
		\label{Fig_02}
	\end{figure*}

	Next, we evaluate the quantum-metric currents in the low-field limit. In this model the intrinsic contribution ($j_{\rm qmi}$) vanishes for both zero and finite tilt, so the low-field response is entirely extrinsic. Finite tilt enhances the magnitude of that extrinsic current without changing this qualitative conclusion. Fig.~\ref{Fig_02}(a) shows the variation of the longitudinal ($j_{\text{qme}}^{||}$) and transverse ($j_{\text{qme}}^{\perp}$) responses with respect to the chemical potential for the same field and Hamiltonian parameters. Within the gapless low-energy model, the quantum-metric current is strongly enhanced as $\mu\to0$. In any realistic system, this growth is cut off by a finite gap, disorder broadening, temperature, or the breakdown of the continuum model. This behavior is distinct from that of the conventional Bloch current, which scales linearly with chemical potential, as shown in Fig.~\ref{Fig_02}(b). For the field configuration used in Fig.~\ref{Fig_02}, the transverse response vanishes because the applied field preserves the mirror symmetry $M_y$. A finite transverse quantum-metric current appears once the field has a nonzero cross-gradient component, such as $E_{xy}=E_{yx}$. 
	Fig.~\ref{Fig_02}(a) shows that the quantum-metric response is largest near the charge-neutrality point. This follows from the quantum-metric scaling $g\sim 1/\epsilon_0^4$, which enhances the contribution from low-energy states. As the chemical potential moves away from charge neutrality, the quantum metric is suppressed and the current correspondingly decreases, directly linking the conductivity trend to the underlying band geometry. In the low-temperature limit, the extrinsic quantum-metric current density at a point $\bm{R}$ is
	\begin{equation}\label{j_qme_anl1}
		j^x_{\text{qme}} (\bm{R},\mu) = -\alpha(\bm R)\tau E_{11} \frac{{3\pi v_t v_F}}{\mu^2}. 
	\end{equation}
	Here $\alpha = \alpha_c$ for $c=x$, the direction of the perturbing field. 
	From Eq.~\eqref{j_qme_anl1}, we see that the current varies as $1/\mu^2$, in clear contrast to the Bloch current, which instead scales as $\mu$. This singular low-energy trend is a property of the gapless continuum approximation. In a gapped realization, where the gap is sufficiently large to suppress Zener tunneling but keeps the Berry curvature negligible, the quantum metric would remain large near the band edge but would be regularized by the gap and the finite bandwidth. Operationally, the divergent scale is cut off by the largest of the gaps, disorder broadening, temperature, or the breakdown scale of the low-energy model. The expected experimental signature is therefore a strong low-density enhancement, not an actual divergence at charge neutrality. The quantum-metric current then vanishes asymptotically at large $\mu$, consistent with Fig.~\ref{Fig_02}(a). 
	
	We have also discussed the isotropic field case ($\alpha_c $ for $c\in \{x,y\}$ and $E_{ij}=E_{11}$ $\forall~i,j\in\{x,y\}$) in Appendix~\ref{T=0 calc}, which shows similar current behavior with the chemical potential. To discuss how the different current components can be separated experimentally, we revisit Eqs.~\eqref{j_qmi} and \eqref{jqme_a}. Reversing the uniform component of the applied field, $\tilde{E}_{c} \rightarrow - \tilde{E}_{c}$, while keeping the imposed gradient tensor fixed leaves $j^a_{\text{qmi}}$ unchanged and reverses the sign of $j^a_{\text{qme}}$. If both the uniform field and the field gradient are reversed together, then $j^a_{\text{qmi}}$ is odd and $j^a_{\text{qme}}$ is even to lowest order. In the present model the intrinsic and extrinsic contributions are therefore sharply distinguishable, and the same parity logic continues to provide an experimental handle even when both responses are of comparable magnitude.
	Thus, the field-reversal parity and scattering-time scaling provide complementary ways to distinguish the different response channels.
	
	Because the spatially varying field generates position-dependent local currents, the measurable current is obtained by averaging over a symmetric interval centered at the origin, $\frac{1}{l}\int_{-l/2}^{l/2}d\bm{R}^{\prime}j(\bm{R}^{\prime})$, where $l$ is the sample length along the current direction. Terms linear in $\bm{R}^{\prime}$ then cancel by symmetry, while the $\bm{R}^{\prime}$-independent contributions remain, leaving a finite net current that scales as $|\bm{E}^{[1]}|E_{11}$.
	
	Summarizing this section, the tilted Dirac model shows that a dominant extrinsic response rooted in band geometry can arise even when the Berry curvature vanishes. Despite remaining a simplified toy model, it respects the underlying symmetry constraints and maintains transparent scaling behavior. Experimental observation of the oscillations requires a system with a sufficiently large unit cell to sustain coherent motion over a timescale exceeding the scattering time, together with a finite band gap large enough to suppress Zener transitions \cite{Khomeriki2016, Parkavi2021, Bouchard1995, Zener1934, Phong2023, Vandenberghe2010}. Flat-band systems with vanishing Berry curvature, finite quantum metric, and an adequate gap are therefore promising candidates for realizing the proposed response \cite{Biesenthal2019, Yoshida2019}. Such platforms would provide a concrete route to probe quantum-metric-driven transport without invoking higher-order band-geometric quantities~\cite{Bhalla2022,Ghorai_prl2025,Mandal_prb2024,Sarkar_PIE2026,Sarkar_njp2025}.
	
	For a simple scale estimate, consider an engineered electronic superlattice or moir\'e flat-band platform~\cite{Adak_NRM2024} with an effective lattice period $a_{\rm eff}\sim 50$--$100$ nm, a bandwidth $W\sim 5$--$20$ meV, an interband gap $\Delta\sim 10$--$50$ meV, and a relaxation time $\tau\sim 1$--$10$ ps. Resolving at least one Bloch period before scattering requires $\omega_B\tau=eEa_{\rm eff}\tau/\hbar \gtrsim 1$, which corresponds to a field strength of order $10^3$--$10^5$ V/m. At the same time, suppressing interband Landau-Zener tunneling requires $eEa_{\rm eff}\ll \Delta^2/W$, giving an upper scale of roughly $10^5$--$10^6$ V/m for the same parameters. A weak inhomogeneity $E_{11}a_{\rm eff}/E\sim 0.05$--$0.1$ then corresponds to gradients of order $10^9$--$10^{11}$ V/m$^2$. These numbers should be read not only as an order-of-magnitude window, but also as an indication that the proposed regime is not parametrically incompatible with such electronic superlattices or artificially engineered optical lattices \cite{Fahimniya2021, Biesenthal2019, Yoshida2019, Longhi2014}.
	
	\section{Conclusion} \label{Section_five}
	
	In summary, we have developed a semiclassical description of Bloch-electron dynamics in weakly inhomogeneous electric fields and identified a band-geometric origin for a real-space oscillation even when the Berry curvature vanishes. The same term also yields intrinsic and extrinsic transport responses, with the extrinsic contribution able to dominate and approach finite high-field saturation when the relative field inhomogeneity is held fixed.
	
	The tilted Dirac example makes the symmetry constraints, density dependence, and field scaling of the response transparent. It also shows, in a concrete setting, that the measured current can be entirely extrinsic. As the model is rather illustrative, the more relevant experimental platforms would be larger-unit-cell systems with a finite quantum metric and a sufficient band gap to suppress Zener tunneling. Flat-band superlattices and optical lattices with the appropriate symmetry structure are natural settings in which to look for these effects \cite{Biesenthal2019,Yoshida2019,Longhi2014}.
	
	Taken together, our results identify an experimentally accessible transport consequence of quantum geometry that does not rely on Berry curvature or on higher-order geometric objects beyond the quantum metric. They provide a route to probing band geometry through inhomogeneous-field dynamics and its strong-field transport signatures.
	
	\section{Acknowledgments}
	We acknowledge financial support from the Science and Engineering Research Board, Government of India, under Project No. MTR/2019/001520, and from the Department of Science and Technology, Government of India, under Project No. DST/NM/TUE/QM-6/2019(G)-IIT Kanpur.
	
	\appendix
	
	\section{The semiclassical formalism} \label{app_sem_cl}
	
	Here, we present a condensed derivation of Eq.~\eqref{net_vel}. In our tensor notation, repeated Roman indices $a,b,c,d$ are summed over. Unless stated otherwise, Greek letters denote spatial components, and a lower index in a denominator is treated as equivalent to an upper index in a numerator, and vice versa. We consider an inhomogeneous time-independent electric field, $\tilde{\bm{E}}$, kept to first order in its gradient \cite{10.1007/3-540-30924-1_7}. The semiclassical wavepacket dynamics are obtained from the velocity and force equations for the first moments of the position and momentum operators, $r^a = \langle \psi(t)| \hat{r}^a| \psi (t) \rangle$ and $k_a = \langle \psi(t)| \hat{q}_a| \psi (t) \rangle$, respectively. The wave function $|\psi (t)\rangle$ is constructed from Bloch eigenstates $|\psi_\textbf{q}\rangle$ as $|\psi(t)\rangle = \int_\textbf{q} n(\textbf{q}, t) |\psi_\textbf{q}\rangle,$ where $\int_\textbf{q} = \int d\textbf{q}/(2\pi)^D$ in $D$ dimensions, and $n(\textbf{q}, t)$ is normalized over the Brillouin zone (BZ). The first and second moments of the position operator within $\psi(t)$ follow from the identities \cite{Lapa2019}
	\begin{eqnarray}
		\langle \hat{r}^a\rangle_{\textbf{q}\textbf{q}'} &=& i \partial_{q_a}\delta(\textbf{q} - \textbf{q}') + \delta(\textbf{q} - \textbf{q}') \mathcal{R}^a (\textbf{q}), \label{x_av}\\  
		\langle \hat{r}^a \hat{r}^b \rangle_{\textbf{q}\textbf{q}'} &=& \partial^2_{q_a q_b} \delta(\textbf{q} - \textbf{q}') - i \mathcal{R}^b \partial_{q'_a} \delta(\textbf{q} - \textbf{q}') \nonumber \\
		&{}& - i \mathcal{R}^a \partial_{q'_b} \delta(\textbf{q} - \textbf{q}') - \delta(\textbf{q} - \textbf{q}')\langle\partial^2_{q_a q_b} \rangle_{\textbf{q}\textbf{q}},\nonumber
		\label{xx_av}\\
	\end{eqnarray}
	where $\langle \cdot \cdot \rangle_{\textbf{q}\textbf{q}'} \equiv \langle \psi_\textbf{q}| \cdot \cdot |\psi_{\textbf{q}'} \rangle,$ and $\mathcal{R}^b = -i\langle\psi_\textbf{q}| \partial_{q_b} \psi_\textbf{q}\rangle$ is the Berry connection \cite{vanderbilt2018berry}. Since the wavepacket is sharply peaked in reciprocal space, we take $|n(\textbf{q}, t)|^2 \approx \delta(\textbf{q} - \textbf{K})$ \cite{PhysRevB.59.14915}. A straightforward calculation from Eqs.~\eqref{x_av} and \eqref{xx_av} yields
	\begin{eqnarray}
		\langle \hat{r}^a \rangle_t &=& r^a  \approx \bigg[\frac{i}{n(\textbf{q},t)}\partial_{q_a} n(\textbf{q}, t)\bigg]_{\textbf{q} = \textbf{k}} + \mathcal{R}^a, \label{first_mmnt_x}\\   
		\langle \hat{r}^a \hat{r}^b \rangle_t &\approx& r^a r^b + g^{a b} (\textbf{k}).\label{second_mmnt_x}
	\end{eqnarray}
	Here $g^{a b}({\bm{k}}) \equiv g^{a b}_{\bm{k}}$ is the Fubini-Study metric \cite{cheng2013quantum, Espinosa-Champo_2024}, and $\langle \cdot \cdot \rangle_t \equiv \langle \psi(t)|\cdot \cdot| \psi (t)\rangle$. The modified semiclassical equations of motion are \cite{Lapa2019}
	\begin{align}
		\Dot{r}^a &= \frac{1}{\hbar}\partial_{k_a} \epsilon_{\bm{k}} - \Omega^{a b}_{\bm{k}}\Dot{k}_b + \frac{e}{2\hbar}E_{b d}\partial_{k_a} g^{b d}_{\bm{k}},\label{x_dot1}\\
		\hbar\Dot{k}_a &= -e\tilde{E}_a ({\bm{r}}),\label{k_dot1}
	\end{align}
	with $\epsilon_{\bm{k}}, \Omega^{a b}_{\bm{k}},$ and $g^{a b}_{\bm{k}}$ denoting the band dispersion, band-resolved Berry curvature, and quantum metric, respectively. The quantum metric and Berry curvature are the real and imaginary parts of the quantum geometric tensor \cite{PhysRevB.104.085114, PhysRevB.81.245129}.
	Here,
	\begin{equation}\label{Eg}
		E_{b d} g^{b d}_{\bm{k}}
		=
		E_{xx} g^{xx}_{\bm{k}} + E_{xy} g^{xy}_{\bm{k}} + E_{yx} g^{yx}_{\bm{k}} + E_{yy} g^{yy}_{\bm{k}}.
	\end{equation}
	For symmetric field-gradient and metric tensors, the off-diagonal terms may equivalently be written as $2E_{xy}g^{xy}_{\bm{k}}$, but no additional factor is absorbed into the definition of the field-gradient tensor.
	Including the electric-field corrections to the energy, the expectation value of the full Hamiltonian is
	\begin{equation}
		\langle \hat{\mathcal{H}}\rangle_t = \langle \hat{\mathcal{H}}_0\rangle_t + eE_a\langle \hat{r}^a \rangle_t +\frac{e}{2} E_{a b}\langle \hat{r}^a \hat{r}^b \rangle_t,\\
	\end{equation}    
	Using Eqs.~\eqref{first_mmnt_x} and \eqref{second_mmnt_x}, this becomes
	\begin{eqnarray} 
		\langle \mathcal{H}\rangle_t &\approx& \epsilon_{\bm{k}} + e E_a r^a  + \frac{e}{2} E_{a b} (r^a r^b + g^{a b}_{\bm{k}}),\\
		&\equiv& \varepsilon_{\text{eff}}. \label{eps_eff}
	\end{eqnarray}
	The gradient field therefore corrects the band energy. The semiclassical equations of motion can be written compactly as
	\begin{eqnarray}
		\Dot{r}^a &=& \frac{1}{\hbar} \frac{\partial \varepsilon_{\text{eff}}}{\partial k_a} - \Omega^{a b}_{\bm{k}}\Dot{k}_b, \label{vel_eq}\\    
		\hbar\Dot{k}_a &=&  -\frac{\partial \varepsilon_{\text{eff}}}{\partial r^a} \label{for_eq}.
	\end{eqnarray}
	Using $\varepsilon_{\text{eff}}$ in Eq.~\eqref{for_eq}, the force equation becomes
	\begin{eqnarray}
		\Dot{k}_a &=& -\frac{e}{\hbar}\bigg[ E_a + \frac{E_{d b}}{2}\partial_{r^a}(r^d r^b)\bigg], \\
		&=& -\frac{e}{\hbar}\bigg[ E_a + \frac{E_{d b}}{2}(r^b \delta^d_a + r^d \delta^b_a)\bigg],\\
		&=& -\frac{e}{\hbar}\bigg[ E_a +\frac{1}{2}(E_{a b}r^b + E_{d a}r^d)\bigg]. \label{k_dot_eq}
	\end{eqnarray}
	Using Eq.~\eqref{k_dot_eq} in Eq.~\eqref{vel_eq} gives
	\begin{equation}\label{full_vel}
		\Dot{r}^a = \frac{1}{\hbar} \frac{\partial \varepsilon_{\text{eff}}}{\partial k_a} + \frac{e}{\hbar} \Omega^{a b}_{\bm{k}}[E_{b} + \frac{1}{2}(E_{b c}r^c + E_{d b}r^d)].
	\end{equation}
	Expanding the first term,
	\begin{eqnarray}
		\frac{1}{\hbar} \frac{\partial \varepsilon_{\text{eff}}}{\partial k_a} &=& \frac{1}{\hbar} \partial_{k_a}[\epsilon_{\bm{k}}  + e E_a r^a + \frac{e}{2} E_{d b} (r^d r^b + g^{d b}_{\bm{k}})], \nonumber\\
		&=& \frac{1}{\hbar} \partial_{k_a}\epsilon_{\bm{k}} + \frac{e}{2\hbar} E_{d b}\partial_{k_a}g^{d b}_{\bm{k}}. \label{eff_vel}
	\end{eqnarray}
	Using Eq.~\eqref{eff_vel} in Eq.~\eqref{full_vel}, and explicitly expanding the terms in $\Dot{r}^a$, gives
	\begin{eqnarray}
		\Dot{r}^a &=& \frac{1}{\hbar} \partial_{k_a}\epsilon_{\bm{k}} + \frac{e}{\hbar} \Omega^{a b}_{\bm{k}} \tilde{E}_{b} + \frac{e}{2\hbar} E_{d b}\partial_{k_a}g^{d b}_{\bm{k}}.
	\end{eqnarray}
	
	\section{Calculations of the oscillations}\label{osc_calc}
	
	In an inhomogeneous electric field $\bm{E}$, kept to first order in its gradient, $\tilde{E}_a = E_a + E_{a b}r^b = \tilde{\bm{E}}(\bm{r}(t))\cdot\hat{r}^a$. The decoupled semiclassical velocity is
	\begin{equation}
		\Dot{r}^a = \frac{1}{\hbar}\partial_{k_a} \epsilon_{\bm{k}} + \frac{e}{\hbar}{\Omega}^{a b}_{\bm{k}}\tilde{E}_b + \frac{e}{2\hbar}E_{b d} \partial_{k_a} g^{b d}_{\bm{k}}.
	\end{equation}
	Expanding the right-hand side in Fourier components converts the momentum-space dependence into the real-space oscillatory response.
	\begin{widetext} 
		\begin{align}
			\Dot{r}^a(\tau') ={} \sum_{E_a a^a_i} &e^{ik_{0a} a^a_i}e^{-ie E_a a^a_i \tau'/\hbar} \bigg[ \frac{1}{\hbar} (ia^a_i) \Tilde{\epsilon}_{\bm{a}_i} + \frac{e}{\hbar}\Tilde{\Omega}^{a b}_{\bm{a}_i}E_b +\frac{e}{2 \hbar}(ia^a_i)\Tilde{g}^{b d}_{\bm{a}_i}E^{(0)}_{b d}\bigg].
		\end{align}
		We now specialize to systems with $\mathcal{P}$, $\mathcal{T}$, or $\mathcal{PT}$ symmetry. Using $\tilde{E}_a = E_a + E_{a b}r^b(t)$,
		\begin{eqnarray}
			\Dot{r}^a(\tau') &=& \sum_{\bm{a}^a_i} e^{ik_{0a} a^a_i}e^{-ie \tilde{E}_a a^a_i \tau'/\hbar} \bigg[ \frac{1}{\hbar}(ia_i^a) \Tilde{\epsilon}_{\bm{a}_i} +\frac{e}{2 \hbar}(ia_i^a)\Tilde{g}^{b d}_{\bm{a}_i}E_{b d}  \bigg], \\
			&\equiv& \sum_{\tilde{E}_a a^a_i} e^{ik_{0a} a^a_i}e^{-ie \tilde{E}_a a^a_i\tau'/\hbar} \bigg[ \frac{1}{\hbar}(ia_i^a) \Tilde{\epsilon}_{\bm{a}_i} + \frac{e}{2 \hbar}(ia_i^a)\Tilde{g}^{b d}_{\bm{a}_i}E_{b d} \bigg]. 
		\end{eqnarray}
	\end{widetext}
	The summation over lattice vectors separates into terms with $\bm{a}_i\cdot \tilde{\bm{E}}=0$ and terms with $\bm{a}_i\cdot \tilde{\bm{E}}\neq 0$. The former produce constant drift, while the latter produce oscillatory motion. Thus,
	\begin{align}
		{\bm{r}}(t) &= \int_0^t d\tau' \bigg[ \sum_{{\bm{a}}_i\cdot \tilde{\bm{E}} =0} \bigg\{ \Dot{{\bm{r}}} \bigg\} +  \sum_{{\bm{a}}_i\cdot\tilde{\bm{E}}\neq 0}\bigg\{ \Dot{{\bm{r}}}\bigg\} \bigg],\\
		&= \underbrace{t \sum_{\substack{{\bm{a}}_i\cdot \tilde{\bm{E}}=0}} \bigg\{ \hspace{.3cm} \bigg\}}_\text{constant drift} 
		+ \underbrace{\int_0^t  d\tau'  \sum_{{\bm{a}}_i\cdot \tilde{\bm{E}}\neq 0}\bigg\{ \hspace{.3cm} \bigg\}}_\text{oscillations}.
		\label{r_of_t}
	\end{align}
	This gives
	\begin{widetext}
		\begin{subequations}
			\begin{eqnarray}
				r^a_{\text{drift}} (t) &=& \frac{t}{\hbar} \sum_{E_c a^c_i = 0} e^{i\bm{k}_{0}\cdot \bm{a}_i} ia^a_i\bigg(\Tilde{\epsilon} + \frac{e}{2} \Tilde{g}^{b d}E_{b d}\bigg), \label{drift}\\
				r^a_{\text{Bloch}}(t) &\simeq&  \sum_{E_c a^c_i \neq 0}e^{i\bm{k}_{0}\cdot \bm{a}_i} [1 - \exp{(-ieE_c a^c_it/\hbar)}]\frac{ a^a_i \Tilde{\epsilon}}{eE_c a^c_i}, \label{bloch_osc}\\
				r^a_{\text{qm}}(t) &\simeq&  \frac{1}{2}\sum_{E_c a^c_i \neq 0} e^{i\bm{k}_{0}\cdot \bm{a}_i} [1 - \exp{(-ie E_c a^c_it/\hbar)}] \frac{ a^a_i\Tilde{g}^{b d}E_{b d} }{E_c a^c_i}. \label{qme_osc}
			\end{eqnarray}
		\end{subequations}
	\end{widetext}
	Within the weak-gradient approximation, Eq.~\eqref{drift} shows that the drift term has both a band-dispersive contribution and a band-geometric contribution from the quantum metric.
	
	\section{Analytical expression of isotropic field responses at zero temperature}\label{T=0 calc}
	
	We calculate the $x$ component of the current density, $\bm{j}_{\text{qme}}$. We have $\epsilon_{\bm{k}} = \eta v_t k_x + \epsilon^{(0)}_{\bm{k}},$ where $\epsilon^{(0)}_{\bm{k}}$ (or $\epsilon_0$) is the dispersion of the conduction band at zero tilt. From our expressions, we have
	\begin{eqnarray*}
		\partial_{k_x} g^\Sigma_{+-} &=& \frac{1}{2} \left(\frac{4 {k_x}^2 {k_y}}{\left({k_x}^2+{k_y}^2\right)^3}-\frac{{k_x}+{k_y}}{\left({k_x}^2+{k_y}^2\right)^2}\right),\\
		(\partial_{k_x} + \partial_{k_y})\epsilon_{\bm{k}} &=& \eta v_t + v_F\frac{(k_x + k_y)}{\sqrt{k^2_x + k^2_y}},
	\end{eqnarray*}
	where $g^\Sigma_{+-} = (g^{xx}_{+-} + g^{yy}_{+-} + g^{xy}_{+-} + g^{yx}_{+-})$. These expressions are inserted into Eq.~\eqref{jqme_a} in the low-temperature limit. We set $E_{ij} = E_{11} ~ \forall i, j \in \{x, y\}$. The integral is then evaluated using the dispersion-azimuth variables $(d\epsilon_0\, d k_\theta)$. Since $\epsilon_0 = v_F \sqrt{k^2_x + k^2_y}$, $dk_x\, dk_y = k_r\, dk_r\, dk_\theta = (1/v_F) \epsilon_0 d\epsilon_0 \, dk_\theta.$ 
	We Taylor expand the Dirac delta function to first derivative order as $\delta(\epsilon - \mu) \simeq \delta(\epsilon_0 - \mu) + \frac{\eta v_t\epsilon_0}{v_F} \cos\theta\frac{\partial}{\partial \epsilon_0}\delta(\epsilon_0 - \mu) $. This gives
	\begin{equation}
		j^x_{\text{qme}} (\mu) = -\frac{{3}}{\mu^2}\left[\frac{\alpha \tau E_{11} v^3_F}{16 \pi} \left(v_F + \frac{\eta^2 v^2_t}{v_F}\right)\right]. \label{j_qme_anl}\\
	\end{equation}
	The same calculation shows that $\bm{j}_{\text{qmi}}$ vanishes at both finite and zero tilt.
	
	\subsection{Bloch currents at zero temperature} \label{Blochcurrent_at_0k}
	Bloch oscillations also give rise to a drift current, $\bm{j}_{\text{Bloch}}$. This current contains intrinsic ($\bm{j}_{\text{Bloch, i}}$) and extrinsic ($\bm{j}_{\text{Bloch, e}}$) contributions. As in the quantum-metric response, the intrinsic component arises from Fermi-sea contributions, while the extrinsic part is associated with Fermi-surface effects. In the low-temperature limit,
	\begin{subequations}
		\begin{eqnarray}
			\bm{j}_{\text{Bloch, e}} &=& \frac{\alpha \tau e}{2 \pi h} [v^2_t + v^2_F \pi + 4 v_t v_F \pi] \mu,\\
			\bm{j}_{\text{Bloch, i}} &=& - \frac{e}{2h} v_t (1 + 2 \eta) \mu^2.
		\end{eqnarray}
	\end{subequations}
	The intrinsic Bloch current is a direct consequence of the tilt and varies quadratically with chemical potential. The extrinsic response depends on both the tilt and the Fermi velocity and varies linearly with chemical potential. Thus, the Bloch currents grow at higher chemical potential, in contrast to the $1/\mu^2$ dependence of the quantum-metric current.
	
	\bibliography{nit3.bib}

\begin{thebibliography}{65}%
\makeatletter
\providecommand \@ifxundefined [1]{%
 \@ifx{#1\undefined}
}%
\providecommand \@ifnum [1]{%
 \ifnum #1\expandafter \@firstoftwo
 \else \expandafter \@secondoftwo
 \fi
}%
\providecommand \@ifx [1]{%
 \ifx #1\expandafter \@firstoftwo
 \else \expandafter \@secondoftwo
 \fi
}%
\providecommand \natexlab [1]{#1}%
\providecommand \enquote  [1]{``#1''}%
\providecommand \bibnamefont  [1]{#1}%
\providecommand \bibfnamefont [1]{#1}%
\providecommand \citenamefont [1]{#1}%
\providecommand \href@noop [0]{\@secondoftwo}%
\providecommand \href [0]{\begingroup \@sanitize@url \@href}%
\providecommand \@href[1]{\@@startlink{#1}\@@href}%
\providecommand \@@href[1]{\endgroup#1\@@endlink}%
\providecommand \@sanitize@url [0]{\catcode `\\12\catcode `\$12\catcode
  `\&12\catcode `\#12\catcode `\^12\catcode `\_12\catcode `\%12\relax}%
\providecommand \@@startlink[1]{}%
\providecommand \@@endlink[0]{}%
\providecommand \url  [0]{\begingroup\@sanitize@url \@url }%
\providecommand \@url [1]{\endgroup\@href {#1}{\urlprefix }}%
\providecommand \urlprefix  [0]{URL }%
\providecommand \Eprint [0]{\href }%
\providecommand \doibase [0]{https://doi.org/}%
\providecommand \selectlanguage [0]{\@gobble}%
\providecommand \bibinfo  [0]{\@secondoftwo}%
\providecommand \bibfield  [0]{\@secondoftwo}%
\providecommand \translation [1]{[#1]}%
\providecommand \BibitemOpen [0]{}%
\providecommand \bibitemStop [0]{}%
\providecommand \bibitemNoStop [0]{.\EOS\space}%
\providecommand \EOS [0]{\spacefactor3000\relax}%
\providecommand \BibitemShut  [1]{\csname bibitem#1\endcsname}%
\let\auto@bib@innerbib\@empty
\bibitem [{\citenamefont {Bloch}(1929)}]{bloch1929}%
  \BibitemOpen
  \bibfield  {author} {\bibinfo {author} {\bibfnamefont {F.}~\bibnamefont
  {Bloch}},\ }\bibfield  {title} {\bibinfo {title} {{\"U}ber die
  quantenmechanik der elektronen in kristallgittern},\ }\href@noop {}
  {\bibfield  {journal} {\bibinfo  {journal} {Zeitschrift f{\"u}r physik}\
  }\textbf {\bibinfo {volume} {52}},\ \bibinfo {pages} {555} (\bibinfo {year}
  {1929})}\BibitemShut {NoStop}%
\bibitem [{\citenamefont {Longhi}(2009)}]{Longhi2009}%
  \BibitemOpen
  \bibfield  {author} {\bibinfo {author} {\bibfnamefont {S.}~\bibnamefont
  {Longhi}},\ }\bibfield  {title} {\bibinfo {title} {Bloch oscillations in
  complex crystals with $\mathcal{P}\mathcal{T}$ symmetry},\ }\href
  {https://doi.org/10.1103/PhysRevLett.103.123601} {\bibfield  {journal}
  {\bibinfo  {journal} {Phys. Rev. Lett.}\ }\textbf {\bibinfo {volume} {103}},\
  \bibinfo {pages} {123601} (\bibinfo {year} {2009})}\BibitemShut {NoStop}%
\bibitem [{\citenamefont {Price}\ and\ \citenamefont
  {Cooper}(2012)}]{Price2012}%
  \BibitemOpen
  \bibfield  {author} {\bibinfo {author} {\bibfnamefont {H.~M.}\ \bibnamefont
  {Price}}\ and\ \bibinfo {author} {\bibfnamefont {N.~R.}\ \bibnamefont
  {Cooper}},\ }\bibfield  {title} {\bibinfo {title} {Mapping the berry
  curvature from semiclassical dynamics in optical lattices},\ }\href
  {https://doi.org/10.1103/PhysRevA.85.033620} {\bibfield  {journal} {\bibinfo
  {journal} {Phys. Rev. A}\ }\textbf {\bibinfo {volume} {85}},\ \bibinfo
  {pages} {033620} (\bibinfo {year} {2012})}\BibitemShut {NoStop}%
\bibitem [{\citenamefont {Gustavsson}\ \emph {et~al.}(2008)\citenamefont
  {Gustavsson}, \citenamefont {Haller}, \citenamefont {Mark}, \citenamefont
  {Danzl}, \citenamefont {Rojas-Kopeinig},\ and\ \citenamefont
  {N\"agerl}}]{Gustavsson2008}%
  \BibitemOpen
  \bibfield  {author} {\bibinfo {author} {\bibfnamefont {M.}~\bibnamefont
  {Gustavsson}}, \bibinfo {author} {\bibfnamefont {E.}~\bibnamefont {Haller}},
  \bibinfo {author} {\bibfnamefont {M.~J.}\ \bibnamefont {Mark}}, \bibinfo
  {author} {\bibfnamefont {J.~G.}\ \bibnamefont {Danzl}}, \bibinfo {author}
  {\bibfnamefont {G.}~\bibnamefont {Rojas-Kopeinig}},\ and\ \bibinfo {author}
  {\bibfnamefont {H.-C.}\ \bibnamefont {N\"agerl}},\ }\bibfield  {title}
  {\bibinfo {title} {Control of interaction-induced dephasing of bloch
  oscillations},\ }\href {https://doi.org/10.1103/PhysRevLett.100.080404}
  {\bibfield  {journal} {\bibinfo  {journal} {Phys. Rev. Lett.}\ }\textbf
  {\bibinfo {volume} {100}},\ \bibinfo {pages} {080404} (\bibinfo {year}
  {2008})}\BibitemShut {NoStop}%
\bibitem [{\citenamefont {Unuma}\ \emph {et~al.}(2006)\citenamefont {Unuma},
  \citenamefont {Sekine},\ and\ \citenamefont {Hirakawa}}]{Unuma2006}%
  \BibitemOpen
  \bibfield  {author} {\bibinfo {author} {\bibfnamefont {T.}~\bibnamefont
  {Unuma}}, \bibinfo {author} {\bibfnamefont {N.}~\bibnamefont {Sekine}},\ and\
  \bibinfo {author} {\bibfnamefont {K.}~\bibnamefont {Hirakawa}},\ }\bibfield
  {title} {\bibinfo {title} {{Dephasing of Bloch oscillating electrons in
  GaAs-based superlattices due to interface roughness scattering}},\ }\href
  {https://doi.org/10.1063/1.2360911} {\bibfield  {journal} {\bibinfo
  {journal} {Applied Physics Letters}\ }\textbf {\bibinfo {volume} {89}},\
  \bibinfo {pages} {161913} (\bibinfo {year} {2006})}\BibitemShut {NoStop}%
\bibitem [{\citenamefont {Meinert}\ \emph {et~al.}(2014)\citenamefont
  {Meinert}, \citenamefont {Mark}, \citenamefont {Kirilov}, \citenamefont
  {Lauber}, \citenamefont {Weinmann}, \citenamefont {Gr\"obner},\ and\
  \citenamefont {N\"agerl}}]{Meinert2014}%
  \BibitemOpen
  \bibfield  {author} {\bibinfo {author} {\bibfnamefont {F.}~\bibnamefont
  {Meinert}}, \bibinfo {author} {\bibfnamefont {M.~J.}\ \bibnamefont {Mark}},
  \bibinfo {author} {\bibfnamefont {E.}~\bibnamefont {Kirilov}}, \bibinfo
  {author} {\bibfnamefont {K.}~\bibnamefont {Lauber}}, \bibinfo {author}
  {\bibfnamefont {P.}~\bibnamefont {Weinmann}}, \bibinfo {author}
  {\bibfnamefont {M.}~\bibnamefont {Gr\"obner}},\ and\ \bibinfo {author}
  {\bibfnamefont {H.-C.}\ \bibnamefont {N\"agerl}},\ }\bibfield  {title}
  {\bibinfo {title} {Interaction-induced quantum phase revivals and evidence
  for the transition to the quantum chaotic regime in 1d atomic bloch
  oscillations},\ }\href {https://doi.org/10.1103/PhysRevLett.112.193003}
  {\bibfield  {journal} {\bibinfo  {journal} {Phys. Rev. Lett.}\ }\textbf
  {\bibinfo {volume} {112}},\ \bibinfo {pages} {193003} (\bibinfo {year}
  {2014})}\BibitemShut {NoStop}%
\bibitem [{\citenamefont {Fahimniya}\ \emph {et~al.}(2021)\citenamefont
  {Fahimniya}, \citenamefont {Dong}, \citenamefont {Kiselev},\ and\
  \citenamefont {Levitov}}]{Fahimniya2021}%
  \BibitemOpen
  \bibfield  {author} {\bibinfo {author} {\bibfnamefont {A.}~\bibnamefont
  {Fahimniya}}, \bibinfo {author} {\bibfnamefont {Z.}~\bibnamefont {Dong}},
  \bibinfo {author} {\bibfnamefont {E.~I.}\ \bibnamefont {Kiselev}},\ and\
  \bibinfo {author} {\bibfnamefont {L.}~\bibnamefont {Levitov}},\ }\bibfield
  {title} {\bibinfo {title} {Synchronizing bloch-oscillating free carriers in
  moir\'e flat bands},\ }\href {https://doi.org/10.1103/PhysRevLett.126.256803}
  {\bibfield  {journal} {\bibinfo  {journal} {Phys. Rev. Lett.}\ }\textbf
  {\bibinfo {volume} {126}},\ \bibinfo {pages} {256803} (\bibinfo {year}
  {2021})}\BibitemShut {NoStop}%
\bibitem [{\citenamefont {Dekorsy}\ \emph {et~al.}(1995)\citenamefont
  {Dekorsy}, \citenamefont {Ott}, \citenamefont {Kurz},\ and\ \citenamefont
  {K\"ohler}}]{Dekorsy1995}%
  \BibitemOpen
  \bibfield  {author} {\bibinfo {author} {\bibfnamefont {T.}~\bibnamefont
  {Dekorsy}}, \bibinfo {author} {\bibfnamefont {R.}~\bibnamefont {Ott}},
  \bibinfo {author} {\bibfnamefont {H.}~\bibnamefont {Kurz}},\ and\ \bibinfo
  {author} {\bibfnamefont {K.}~\bibnamefont {K\"ohler}},\ }\bibfield  {title}
  {\bibinfo {title} {Bloch oscillations at room temperature},\ }\href
  {https://doi.org/10.1103/PhysRevB.51.17275} {\bibfield  {journal} {\bibinfo
  {journal} {Phys. Rev. B}\ }\textbf {\bibinfo {volume} {51}},\ \bibinfo
  {pages} {17275} (\bibinfo {year} {1995})}\BibitemShut {NoStop}%
\bibitem [{\citenamefont {Krieger}\ and\ \citenamefont
  {Iafrate}(1986)}]{Krieger1986}%
  \BibitemOpen
  \bibfield  {author} {\bibinfo {author} {\bibfnamefont {J.~B.}\ \bibnamefont
  {Krieger}}\ and\ \bibinfo {author} {\bibfnamefont {G.~J.}\ \bibnamefont
  {Iafrate}},\ }\bibfield  {title} {\bibinfo {title} {Time evolution of bloch
  electrons in a homogeneous electric field},\ }\href
  {https://doi.org/10.1103/PhysRevB.33.5494} {\bibfield  {journal} {\bibinfo
  {journal} {Phys. Rev. B}\ }\textbf {\bibinfo {volume} {33}},\ \bibinfo
  {pages} {5494} (\bibinfo {year} {1986})}\BibitemShut {NoStop}%
\bibitem [{\citenamefont {Tsu}\ and\ \citenamefont {Esaki}(1973)}]{tsu1973}%
  \BibitemOpen
  \bibfield  {author} {\bibinfo {author} {\bibfnamefont {R.}~\bibnamefont
  {Tsu}}\ and\ \bibinfo {author} {\bibfnamefont {L.}~\bibnamefont {Esaki}},\
  }\bibfield  {title} {\bibinfo {title} {Tunneling in a finite superlattice},\
  }\href@noop {} {\bibfield  {journal} {\bibinfo  {journal} {Applied Physics
  Letters}\ }\textbf {\bibinfo {volume} {22}},\ \bibinfo {pages} {562}
  (\bibinfo {year} {1973})}\BibitemShut {NoStop}%
\bibitem [{\citenamefont {Averin}\ \emph {et~al.}(1985)\citenamefont {Averin},
  \citenamefont {Zorin},\ and\ \citenamefont {Likharev}}]{averin1985}%
  \BibitemOpen
  \bibfield  {author} {\bibinfo {author} {\bibfnamefont {D.}~\bibnamefont
  {Averin}}, \bibinfo {author} {\bibfnamefont {A.}~\bibnamefont {Zorin}},\ and\
  \bibinfo {author} {\bibfnamefont {K.}~\bibnamefont {Likharev}},\ }\bibfield
  {title} {\bibinfo {title} {Bloch oscillations in small josephson junctions},\
  }\href@noop {} {\bibfield  {journal} {\bibinfo  {journal} {Sov. Phys. JETP}\
  }\textbf {\bibinfo {volume} {61}},\ \bibinfo {pages} {407} (\bibinfo {year}
  {1985})}\BibitemShut {NoStop}%
\bibitem [{\citenamefont {Di~Liberto}\ \emph {et~al.}(2020)\citenamefont
  {Di~Liberto}, \citenamefont {Goldman},\ and\ \citenamefont
  {Palumbo}}]{DiLiberto2020}%
  \BibitemOpen
  \bibfield  {author} {\bibinfo {author} {\bibfnamefont {M.}~\bibnamefont
  {Di~Liberto}}, \bibinfo {author} {\bibfnamefont {N.}~\bibnamefont
  {Goldman}},\ and\ \bibinfo {author} {\bibfnamefont {G.}~\bibnamefont
  {Palumbo}},\ }\bibfield  {title} {\bibinfo {title} {Non-abelian bloch
  oscillations in higher-order topological insulators},\ }\href
  {https://doi.org/10.1038/s41467-020-19518-x} {\bibfield  {journal} {\bibinfo
  {journal} {Nature Communications}\ }\textbf {\bibinfo {volume} {11}},\
  \bibinfo {pages} {5942} (\bibinfo {year} {2020})}\BibitemShut {NoStop}%
\bibitem [{\citenamefont {Meinert}\ \emph {et~al.}(2017)\citenamefont
  {Meinert}, \citenamefont {Knap}, \citenamefont {Kirilov}, \citenamefont
  {Jag-Lauber}, \citenamefont {Zvonarev}, \citenamefont {Demler},\ and\
  \citenamefont {Nägerl}}]{Meinert2017}%
  \BibitemOpen
  \bibfield  {author} {\bibinfo {author} {\bibfnamefont {F.}~\bibnamefont
  {Meinert}}, \bibinfo {author} {\bibfnamefont {M.}~\bibnamefont {Knap}},
  \bibinfo {author} {\bibfnamefont {E.}~\bibnamefont {Kirilov}}, \bibinfo
  {author} {\bibfnamefont {K.}~\bibnamefont {Jag-Lauber}}, \bibinfo {author}
  {\bibfnamefont {M.~B.}\ \bibnamefont {Zvonarev}}, \bibinfo {author}
  {\bibfnamefont {E.}~\bibnamefont {Demler}},\ and\ \bibinfo {author}
  {\bibfnamefont {H.-C.}\ \bibnamefont {Nägerl}},\ }\bibfield  {title}
  {\bibinfo {title} {Bloch oscillations in the absence of a lattice},\ }\href
  {https://doi.org/10.1126/science.aah6616} {\bibfield  {journal} {\bibinfo
  {journal} {Science}\ }\textbf {\bibinfo {volume} {356}},\ \bibinfo {pages}
  {945} (\bibinfo {year} {2017})}\BibitemShut {NoStop}%
\bibitem [{\citenamefont {Preiss}\ \emph {et~al.}(2015)\citenamefont {Preiss},
  \citenamefont {Ma}, \citenamefont {Tai}, \citenamefont {Lukin}, \citenamefont
  {Rispoli}, \citenamefont {Zupancic}, \citenamefont {Lahini}, \citenamefont
  {Islam},\ and\ \citenamefont {Greiner}}]{Preiss2015}%
  \BibitemOpen
  \bibfield  {author} {\bibinfo {author} {\bibfnamefont {P.~M.}\ \bibnamefont
  {Preiss}}, \bibinfo {author} {\bibfnamefont {R.}~\bibnamefont {Ma}}, \bibinfo
  {author} {\bibfnamefont {M.~E.}\ \bibnamefont {Tai}}, \bibinfo {author}
  {\bibfnamefont {A.}~\bibnamefont {Lukin}}, \bibinfo {author} {\bibfnamefont
  {M.}~\bibnamefont {Rispoli}}, \bibinfo {author} {\bibfnamefont
  {P.}~\bibnamefont {Zupancic}}, \bibinfo {author} {\bibfnamefont
  {Y.}~\bibnamefont {Lahini}}, \bibinfo {author} {\bibfnamefont
  {R.}~\bibnamefont {Islam}},\ and\ \bibinfo {author} {\bibfnamefont
  {M.}~\bibnamefont {Greiner}},\ }\bibfield  {title} {\bibinfo {title}
  {Strongly correlated quantum walks in optical lattices},\ }\href
  {https://doi.org/10.1126/science.1260364} {\bibfield  {journal} {\bibinfo
  {journal} {Science}\ }\textbf {\bibinfo {volume} {347}},\ \bibinfo {pages}
  {1229} (\bibinfo {year} {2015})}\BibitemShut {NoStop}%
\bibitem [{\citenamefont {Qin}\ \emph {et~al.}(2024)\citenamefont {Qin},
  \citenamefont {Zhang},\ and\ \citenamefont {Li}}]{Qin2024}%
  \BibitemOpen
  \bibfield  {author} {\bibinfo {author} {\bibfnamefont {Y.}~\bibnamefont
  {Qin}}, \bibinfo {author} {\bibfnamefont {K.}~\bibnamefont {Zhang}},\ and\
  \bibinfo {author} {\bibfnamefont {L.}~\bibnamefont {Li}},\ }\bibfield
  {title} {\bibinfo {title} {Geometry-dependent skin effect and anisotropic
  bloch oscillations in a non-hermitian optical lattice},\ }\href
  {https://doi.org/10.1103/PhysRevA.109.023317} {\bibfield  {journal} {\bibinfo
   {journal} {Phys. Rev. A}\ }\textbf {\bibinfo {volume} {109}},\ \bibinfo
  {pages} {023317} (\bibinfo {year} {2024})}\BibitemShut {NoStop}%
\bibitem [{\citenamefont {Pagel}\ \emph {et~al.}(2020)\citenamefont {Pagel},
  \citenamefont {Zhong}, \citenamefont {Parker}, \citenamefont {Olund},
  \citenamefont {Yao},\ and\ \citenamefont {M\"uller}}]{Pagel2020}%
  \BibitemOpen
  \bibfield  {author} {\bibinfo {author} {\bibfnamefont {Z.}~\bibnamefont
  {Pagel}}, \bibinfo {author} {\bibfnamefont {W.}~\bibnamefont {Zhong}},
  \bibinfo {author} {\bibfnamefont {R.~H.}\ \bibnamefont {Parker}}, \bibinfo
  {author} {\bibfnamefont {C.~T.}\ \bibnamefont {Olund}}, \bibinfo {author}
  {\bibfnamefont {N.~Y.}\ \bibnamefont {Yao}},\ and\ \bibinfo {author}
  {\bibfnamefont {H.}~\bibnamefont {M\"uller}},\ }\bibfield  {title} {\bibinfo
  {title} {Symmetric bloch oscillations of matter waves},\ }\href
  {https://doi.org/10.1103/PhysRevA.102.053312} {\bibfield  {journal} {\bibinfo
   {journal} {Phys. Rev. A}\ }\textbf {\bibinfo {volume} {102}},\ \bibinfo
  {pages} {053312} (\bibinfo {year} {2020})}\BibitemShut {NoStop}%
\bibitem [{\citenamefont {Lebugle}\ \emph {et~al.}(2015)\citenamefont
  {Lebugle}, \citenamefont {Gr{\"a}fe}, \citenamefont {Heilmann}, \citenamefont
  {Perez-Leija}, \citenamefont {Nolte},\ and\ \citenamefont
  {Szameit}}]{Lebugle2015}%
  \BibitemOpen
  \bibfield  {author} {\bibinfo {author} {\bibfnamefont {M.}~\bibnamefont
  {Lebugle}}, \bibinfo {author} {\bibfnamefont {M.}~\bibnamefont {Gr{\"a}fe}},
  \bibinfo {author} {\bibfnamefont {R.}~\bibnamefont {Heilmann}}, \bibinfo
  {author} {\bibfnamefont {A.}~\bibnamefont {Perez-Leija}}, \bibinfo {author}
  {\bibfnamefont {S.}~\bibnamefont {Nolte}},\ and\ \bibinfo {author}
  {\bibfnamefont {A.}~\bibnamefont {Szameit}},\ }\bibfield  {title} {\bibinfo
  {title} {Experimental observation of n00n state bloch oscillations},\ }\href
  {https://doi.org/10.1038/ncomms9273} {\bibfield  {journal} {\bibinfo
  {journal} {Nature Communications}\ }\textbf {\bibinfo {volume} {6}},\
  \bibinfo {pages} {8273} (\bibinfo {year} {2015})}\BibitemShut {NoStop}%
\bibitem [{\citenamefont {Corrielli}\ \emph {et~al.}(2013)\citenamefont
  {Corrielli}, \citenamefont {Crespi}, \citenamefont {Della~Valle},
  \citenamefont {Longhi},\ and\ \citenamefont {Osellame}}]{Corrielli2013}%
  \BibitemOpen
  \bibfield  {author} {\bibinfo {author} {\bibfnamefont {G.}~\bibnamefont
  {Corrielli}}, \bibinfo {author} {\bibfnamefont {A.}~\bibnamefont {Crespi}},
  \bibinfo {author} {\bibfnamefont {G.}~\bibnamefont {Della~Valle}}, \bibinfo
  {author} {\bibfnamefont {S.}~\bibnamefont {Longhi}},\ and\ \bibinfo {author}
  {\bibfnamefont {R.}~\bibnamefont {Osellame}},\ }\bibfield  {title} {\bibinfo
  {title} {Fractional bloch oscillations in photonic lattices},\ }\href
  {https://doi.org/10.1038/ncomms2578} {\bibfield  {journal} {\bibinfo
  {journal} {Nature Communications}\ }\textbf {\bibinfo {volume} {4}},\
  \bibinfo {pages} {1555} (\bibinfo {year} {2013})}\BibitemShut {NoStop}%
\bibitem [{\citenamefont {Xu}\ \emph {et~al.}(2016)\citenamefont {Xu},
  \citenamefont {Fegadolli}, \citenamefont {Gan}, \citenamefont {Lu},
  \citenamefont {Liu}, \citenamefont {Li}, \citenamefont {Scherer},\ and\
  \citenamefont {Chen}}]{Xu2016}%
  \BibitemOpen
  \bibfield  {author} {\bibinfo {author} {\bibfnamefont {Y.-L.}\ \bibnamefont
  {Xu}}, \bibinfo {author} {\bibfnamefont {W.~S.}\ \bibnamefont {Fegadolli}},
  \bibinfo {author} {\bibfnamefont {L.}~\bibnamefont {Gan}}, \bibinfo {author}
  {\bibfnamefont {M.-H.}\ \bibnamefont {Lu}}, \bibinfo {author} {\bibfnamefont
  {X.-P.}\ \bibnamefont {Liu}}, \bibinfo {author} {\bibfnamefont {Z.-Y.}\
  \bibnamefont {Li}}, \bibinfo {author} {\bibfnamefont {A.}~\bibnamefont
  {Scherer}},\ and\ \bibinfo {author} {\bibfnamefont {Y.-F.}\ \bibnamefont
  {Chen}},\ }\bibfield  {title} {\bibinfo {title} {Experimental realization of
  bloch oscillations in a parity-time synthetic silicon photonic lattice},\
  }\href {https://doi.org/10.1038/ncomms11319} {\bibfield  {journal} {\bibinfo
  {journal} {Nature Communications}\ }\textbf {\bibinfo {volume} {7}},\
  \bibinfo {pages} {11319} (\bibinfo {year} {2016})}\BibitemShut {NoStop}%
\bibitem [{\citenamefont {Floss}\ \emph {et~al.}(2015)\citenamefont {Floss},
  \citenamefont {Kamalov}, \citenamefont {Averbukh},\ and\ \citenamefont
  {Bucksbaum}}]{Floss2015}%
  \BibitemOpen
  \bibfield  {author} {\bibinfo {author} {\bibfnamefont {J.}~\bibnamefont
  {Floss}}, \bibinfo {author} {\bibfnamefont {A.}~\bibnamefont {Kamalov}},
  \bibinfo {author} {\bibfnamefont {I.~S.}\ \bibnamefont {Averbukh}},\ and\
  \bibinfo {author} {\bibfnamefont {P.~H.}\ \bibnamefont {Bucksbaum}},\
  }\bibfield  {title} {\bibinfo {title} {Observation of bloch oscillations in
  molecular rotation},\ }\href {https://doi.org/10.1103/PhysRevLett.115.203002}
  {\bibfield  {journal} {\bibinfo  {journal} {Phys. Rev. Lett.}\ }\textbf
  {\bibinfo {volume} {115}},\ \bibinfo {pages} {203002} (\bibinfo {year}
  {2015})}\BibitemShut {NoStop}%
\bibitem [{\citenamefont {Chen}\ \emph {et~al.}(2021)\citenamefont {Chen},
  \citenamefont {Yang}, \citenamefont {Qin}, \citenamefont {Li}, \citenamefont
  {Wang}, \citenamefont {Han}, \citenamefont {Zhang}, \citenamefont {Liu},
  \citenamefont {Wang}, \citenamefont {Long}, \citenamefont {Zhang},\ and\
  \citenamefont {Lu}}]{Chen2021}%
  \BibitemOpen
  \bibfield  {author} {\bibinfo {author} {\bibfnamefont {H.}~\bibnamefont
  {Chen}}, \bibinfo {author} {\bibfnamefont {N.}~\bibnamefont {Yang}}, \bibinfo
  {author} {\bibfnamefont {C.}~\bibnamefont {Qin}}, \bibinfo {author}
  {\bibfnamefont {W.}~\bibnamefont {Li}}, \bibinfo {author} {\bibfnamefont
  {B.}~\bibnamefont {Wang}}, \bibinfo {author} {\bibfnamefont {T.}~\bibnamefont
  {Han}}, \bibinfo {author} {\bibfnamefont {C.}~\bibnamefont {Zhang}}, \bibinfo
  {author} {\bibfnamefont {W.}~\bibnamefont {Liu}}, \bibinfo {author}
  {\bibfnamefont {K.}~\bibnamefont {Wang}}, \bibinfo {author} {\bibfnamefont
  {H.}~\bibnamefont {Long}}, \bibinfo {author} {\bibfnamefont {X.}~\bibnamefont
  {Zhang}},\ and\ \bibinfo {author} {\bibfnamefont {P.}~\bibnamefont {Lu}},\
  }\bibfield  {title} {\bibinfo {title} {Real-time observation of frequency
  bloch oscillations with fibre loop modulation},\ }\href
  {https://doi.org/10.1038/s41377-021-00494-w} {\bibfield  {journal} {\bibinfo
  {journal} {Light: Science {\&} Applications}\ }\textbf {\bibinfo {volume}
  {10}},\ \bibinfo {pages} {48} (\bibinfo {year} {2021})}\BibitemShut {NoStop}%
\bibitem [{\citenamefont {Merlin}(2024)}]{Merlin2024}%
  \BibitemOpen
  \bibfield  {author} {\bibinfo {author} {\bibfnamefont {R.}~\bibnamefont
  {Merlin}},\ }\bibfield  {title} {\bibinfo {title} {Phonon bloch
  oscillations},\ }\href {https://doi.org/10.1103/PhysRevLett.132.246302}
  {\bibfield  {journal} {\bibinfo  {journal} {Phys. Rev. Lett.}\ }\textbf
  {\bibinfo {volume} {132}},\ \bibinfo {pages} {246302} (\bibinfo {year}
  {2024})}\BibitemShut {NoStop}%
\bibitem [{\citenamefont {Phong}\ and\ \citenamefont {Mele}(2023)}]{Phong2023}%
  \BibitemOpen
  \bibfield  {author} {\bibinfo {author} {\bibfnamefont {V.~o.~T.}\
  \bibnamefont {Phong}}\ and\ \bibinfo {author} {\bibfnamefont {E.~J.}\
  \bibnamefont {Mele}},\ }\bibfield  {title} {\bibinfo {title} {Quantum
  geometric oscillations in two-dimensional flat-band solids},\ }\href
  {https://doi.org/10.1103/PhysRevLett.130.266601} {\bibfield  {journal}
  {\bibinfo  {journal} {Phys. Rev. Lett.}\ }\textbf {\bibinfo {volume} {130}},\
  \bibinfo {pages} {266601} (\bibinfo {year} {2023})}\BibitemShut {NoStop}%
\bibitem [{\citenamefont {De~Beule}\ and\ \citenamefont
  {Mele}(2023)}]{Beule2023}%
  \BibitemOpen
  \bibfield  {author} {\bibinfo {author} {\bibfnamefont {C.}~\bibnamefont
  {De~Beule}}\ and\ \bibinfo {author} {\bibfnamefont {E.~J.}\ \bibnamefont
  {Mele}},\ }\bibfield  {title} {\bibinfo {title} {Berry curvature spectroscopy
  from bloch oscillations},\ }\href
  {https://doi.org/10.1103/PhysRevLett.131.196603} {\bibfield  {journal}
  {\bibinfo  {journal} {Phys. Rev. Lett.}\ }\textbf {\bibinfo {volume} {131}},\
  \bibinfo {pages} {196603} (\bibinfo {year} {2023})}\BibitemShut {NoStop}%
\bibitem [{\citenamefont {Davies}\ \emph {et~al.}(2004)\citenamefont {Davies},
  \citenamefont {Linfield}, \citenamefont {Pepper},\ and\ \citenamefont
  {Chamberlain}}]{Davies2004}%
  \BibitemOpen
  \bibfield  {author} {\bibinfo {author} {\bibfnamefont {A.~G.}\ \bibnamefont
  {Davies}}, \bibinfo {author} {\bibfnamefont {E.~H.}\ \bibnamefont
  {Linfield}}, \bibinfo {author} {\bibfnamefont {M.}~\bibnamefont {Pepper}},\
  and\ \bibinfo {author} {\bibfnamefont {J.~M.}\ \bibnamefont {Chamberlain}},\
  }\bibfield  {title} {\bibinfo {title} {Where optics meets electronics: recent
  progress in decreasing the terahertz gap},\ }\href
  {https://doi.org/10.1098/rsta.2003.1312} {\bibfield  {journal} {\bibinfo
  {journal} {Philosophical Transactions of the Royal Society of London. Series
  A: Mathematical, Physical and Engineering Sciences}\ }\textbf {\bibinfo
  {volume} {362}},\ \bibinfo {pages} {199} (\bibinfo {year}
  {2004})}\BibitemShut {NoStop}%
\bibitem [{\citenamefont {Borak}(2005)}]{Borak2005}%
  \BibitemOpen
  \bibfield  {author} {\bibinfo {author} {\bibfnamefont {A.}~\bibnamefont
  {Borak}},\ }\bibfield  {title} {\bibinfo {title} {Toward bridging the
  terahertz gap with silicon-based lasers},\ }\href
  {https://doi.org/10.1126/science.1109831} {\bibfield  {journal} {\bibinfo
  {journal} {Science}\ }\textbf {\bibinfo {volume} {308}},\ \bibinfo {pages}
  {638} (\bibinfo {year} {2005})}\BibitemShut {NoStop}%
\bibitem [{\citenamefont {Zeng}\ \emph {et~al.}(2025)\citenamefont {Zeng},
  \citenamefont {Zhou}, \citenamefont {Duan},\ and\ \citenamefont
  {Huang}}]{ZengPRB2025}%
  \BibitemOpen
  \bibfield  {author} {\bibinfo {author} {\bibfnamefont {H.}~\bibnamefont
  {Zeng}}, \bibinfo {author} {\bibfnamefont {Z.}~\bibnamefont {Zhou}}, \bibinfo
  {author} {\bibfnamefont {W.}~\bibnamefont {Duan}},\ and\ \bibinfo {author}
  {\bibfnamefont {H.}~\bibnamefont {Huang}},\ }\bibfield  {title} {\bibinfo
  {title} {Quantum metric induced oscillations in nearly dispersionless flat
  bands},\ }\href {https://doi.org/10.1103/PhysRevB.111.L121102} {\bibfield
  {journal} {\bibinfo  {journal} {Phys. Rev. B}\ }\textbf {\bibinfo {volume}
  {111}},\ \bibinfo {pages} {L121102} (\bibinfo {year} {2025})}\BibitemShut
  {NoStop}%
\bibitem [{\citenamefont {Lapa}\ and\ \citenamefont {Hughes}(2019)}]{Lapa2019}%
  \BibitemOpen
  \bibfield  {author} {\bibinfo {author} {\bibfnamefont {M.~F.}\ \bibnamefont
  {Lapa}}\ and\ \bibinfo {author} {\bibfnamefont {T.~L.}\ \bibnamefont
  {Hughes}},\ }\bibfield  {title} {\bibinfo {title} {Semiclassical wave packet
  dynamics in nonuniform electric fields},\ }\href
  {https://doi.org/10.1103/PhysRevB.99.121111} {\bibfield  {journal} {\bibinfo
  {journal} {Phys. Rev. B}\ }\textbf {\bibinfo {volume} {99}},\ \bibinfo
  {pages} {121111} (\bibinfo {year} {2019})}\BibitemShut {NoStop}%
\bibitem [{\citenamefont {Ashcroft}\ and\ \citenamefont
  {Mermin}(1976)}]{ashcroft1976}%
  \BibitemOpen
  \bibfield  {author} {\bibinfo {author} {\bibfnamefont {N.}~\bibnamefont
  {Ashcroft}}\ and\ \bibinfo {author} {\bibfnamefont {N.}~\bibnamefont
  {Mermin}},\ }\href {https://books.google.co.in/books?id=oXIfAQAAMAAJ} {\emph
  {\bibinfo {title} {Solid State Physics}}},\ HRW international editions\
  (\bibinfo  {publisher} {Holt, Rinehart and Winston},\ \bibinfo {year}
  {1976})\BibitemShut {NoStop}%
\bibitem [{\citenamefont {Aversa}\ and\ \citenamefont
  {Sipe}(1995)}]{Aversa1995}%
  \BibitemOpen
  \bibfield  {author} {\bibinfo {author} {\bibfnamefont {C.}~\bibnamefont
  {Aversa}}\ and\ \bibinfo {author} {\bibfnamefont {J.~E.}\ \bibnamefont
  {Sipe}},\ }\bibfield  {title} {\bibinfo {title} {Nonlinear optical
  susceptibilities of semiconductors: Results with a length-gauge analysis},\
  }\href {https://doi.org/10.1103/PhysRevB.52.14636} {\bibfield  {journal}
  {\bibinfo  {journal} {Phys. Rev. B}\ }\textbf {\bibinfo {volume} {52}},\
  \bibinfo {pages} {14636} (\bibinfo {year} {1995})}\BibitemShut {NoStop}%
\bibitem [{\citenamefont {Kumar}\ \emph {et~al.}(2024)\citenamefont {Kumar},
  \citenamefont {Sarkar},\ and\ \citenamefont {Agarwal}}]{Kumar2024}%
  \BibitemOpen
  \bibfield  {author} {\bibinfo {author} {\bibfnamefont {M.~M.}\ \bibnamefont
  {Kumar}}, \bibinfo {author} {\bibfnamefont {S.}~\bibnamefont {Sarkar}},\ and\
  \bibinfo {author} {\bibfnamefont {A.}~\bibnamefont {Agarwal}},\ }\bibfield
  {title} {\bibinfo {title} {Band geometry induced electro-optic effect and
  polarization rotation},\ }\href {https://doi.org/10.1103/PhysRevB.110.125401}
  {\bibfield  {journal} {\bibinfo  {journal} {Phys. Rev. B}\ }\textbf {\bibinfo
  {volume} {110}},\ \bibinfo {pages} {125401} (\bibinfo {year}
  {2024})}\BibitemShut {NoStop}%
\bibitem [{\citenamefont {Kumar}\ \emph {et~al.}(2025)\citenamefont {Kumar},
  \citenamefont {Sarkar},\ and\ \citenamefont {Agarwal}}]{kumar2025}%
  \BibitemOpen
  \bibfield  {author} {\bibinfo {author} {\bibfnamefont {M.~M.}\ \bibnamefont
  {Kumar}}, \bibinfo {author} {\bibfnamefont {S.}~\bibnamefont {Sarkar}},\ and\
  \bibinfo {author} {\bibfnamefont {A.}~\bibnamefont {Agarwal}},\ }\href
  {https://arxiv.org/abs/2509.13242} {\bibinfo {title} {Band geometric
  transverse current driven by inhomogeneous ac electric field}} (\bibinfo
  {year} {2025}),\ \Eprint {https://arxiv.org/abs/2509.13242} {arXiv:2509.13242
  [cond-mat.mes-hall]} \BibitemShut {NoStop}%
\bibitem [{\citenamefont {Bonilla}\ \emph {et~al.}(2011)\citenamefont
  {Bonilla}, \citenamefont {\'Alvaro},\ and\ \citenamefont
  {Carretero}}]{Bonilla2011}%
  \BibitemOpen
  \bibfield  {author} {\bibinfo {author} {\bibfnamefont {L.~L.}\ \bibnamefont
  {Bonilla}}, \bibinfo {author} {\bibfnamefont {M.}~\bibnamefont {\'Alvaro}},\
  and\ \bibinfo {author} {\bibfnamefont {M.}~\bibnamefont {Carretero}},\
  }\bibfield  {title} {\bibinfo {title} {Theory of spatially inhomogeneous
  bloch oscillations in semiconductor superlattices},\ }\href
  {https://doi.org/10.1103/PhysRevB.84.155316} {\bibfield  {journal} {\bibinfo
  {journal} {Phys. Rev. B}\ }\textbf {\bibinfo {volume} {84}},\ \bibinfo
  {pages} {155316} (\bibinfo {year} {2011})}\BibitemShut {NoStop}%
\bibitem [{\citenamefont {Bhalla}\ \emph {et~al.}(2022)\citenamefont {Bhalla},
  \citenamefont {Das}, \citenamefont {Culcer},\ and\ \citenamefont
  {Agarwal}}]{Bhalla2022}%
  \BibitemOpen
  \bibfield  {author} {\bibinfo {author} {\bibfnamefont {P.}~\bibnamefont
  {Bhalla}}, \bibinfo {author} {\bibfnamefont {K.}~\bibnamefont {Das}},
  \bibinfo {author} {\bibfnamefont {D.}~\bibnamefont {Culcer}},\ and\ \bibinfo
  {author} {\bibfnamefont {A.}~\bibnamefont {Agarwal}},\ }\bibfield  {title}
  {\bibinfo {title} {Resonant second-harmonic generation as a probe of quantum
  geometry},\ }\href {https://doi.org/10.1103/PhysRevLett.129.227401}
  {\bibfield  {journal} {\bibinfo  {journal} {Phys. Rev. Lett.}\ }\textbf
  {\bibinfo {volume} {129}},\ \bibinfo {pages} {227401} (\bibinfo {year}
  {2022})}\BibitemShut {NoStop}%
\bibitem [{\citenamefont {Varshney}\ \emph {et~al.}(2023)\citenamefont
  {Varshney}, \citenamefont {Das}, \citenamefont {Bhalla},\ and\ \citenamefont
  {Agarwal}}]{Varshney2023}%
  \BibitemOpen
  \bibfield  {author} {\bibinfo {author} {\bibfnamefont {H.}~\bibnamefont
  {Varshney}}, \bibinfo {author} {\bibfnamefont {K.}~\bibnamefont {Das}},
  \bibinfo {author} {\bibfnamefont {P.}~\bibnamefont {Bhalla}},\ and\ \bibinfo
  {author} {\bibfnamefont {A.}~\bibnamefont {Agarwal}},\ }\bibfield  {title}
  {\bibinfo {title} {Quantum kinetic theory of nonlinear thermal current},\
  }\href {https://doi.org/10.1103/PhysRevB.107.235419} {\bibfield  {journal}
  {\bibinfo  {journal} {Phys. Rev. B}\ }\textbf {\bibinfo {volume} {107}},\
  \bibinfo {pages} {235419} (\bibinfo {year} {2023})}\BibitemShut {NoStop}%
\bibitem [{\citenamefont {Chang}\ and\ \citenamefont {Niu}(1995)}]{Chang1995}%
  \BibitemOpen
  \bibfield  {author} {\bibinfo {author} {\bibfnamefont {M.-C.}\ \bibnamefont
  {Chang}}\ and\ \bibinfo {author} {\bibfnamefont {Q.}~\bibnamefont {Niu}},\
  }\bibfield  {title} {\bibinfo {title} {Berry phase, hyperorbits, and the
  hofstadter spectrum},\ }\href {https://doi.org/10.1103/PhysRevLett.75.1348}
  {\bibfield  {journal} {\bibinfo  {journal} {Phys. Rev. Lett.}\ }\textbf
  {\bibinfo {volume} {75}},\ \bibinfo {pages} {1348} (\bibinfo {year}
  {1995})}\BibitemShut {NoStop}%
\bibitem [{\citenamefont {Das}\ and\ \citenamefont {Agarwal}(2019)}]{Das2019}%
  \BibitemOpen
  \bibfield  {author} {\bibinfo {author} {\bibfnamefont {K.}~\bibnamefont
  {Das}}\ and\ \bibinfo {author} {\bibfnamefont {A.}~\bibnamefont {Agarwal}},\
  }\bibfield  {title} {\bibinfo {title} {Linear magnetochiral transport in
  tilted type-i and type-ii weyl semimetals},\ }\href
  {https://doi.org/10.1103/PhysRevB.99.085405} {\bibfield  {journal} {\bibinfo
  {journal} {Phys. Rev. B}\ }\textbf {\bibinfo {volume} {99}},\ \bibinfo
  {pages} {085405} (\bibinfo {year} {2019})}\BibitemShut {NoStop}%
\bibitem [{\citenamefont {Mazumder}(2021)}]{mazumder2021}%
  \BibitemOpen
  \bibfield  {author} {\bibinfo {author} {\bibfnamefont {S.}~\bibnamefont
  {Mazumder}},\ }\bibfield  {title} {\bibinfo {title} {Boltzmann transport
  equation based modeling of phonon heat conduction: progress and challenges},\
  }\href {https://doi.org/10.1615/AnnualRevHeatTransfer.2022041316} {\bibfield
  {journal} {\bibinfo  {journal} {Annual Review of Heat Transfer}\ }\textbf
  {\bibinfo {volume} {24}},\ \bibinfo {pages} {71} (\bibinfo {year}
  {2021})}\BibitemShut {NoStop}%
\bibitem [{\citenamefont {Sinitsyn}\ \emph {et~al.}(2005)\citenamefont
  {Sinitsyn}, \citenamefont {Niu}, \citenamefont {Sinova},\ and\ \citenamefont
  {Nomura}}]{Sinitsyn2005}%
  \BibitemOpen
  \bibfield  {author} {\bibinfo {author} {\bibfnamefont {N.~A.}\ \bibnamefont
  {Sinitsyn}}, \bibinfo {author} {\bibfnamefont {Q.}~\bibnamefont {Niu}},
  \bibinfo {author} {\bibfnamefont {J.}~\bibnamefont {Sinova}},\ and\ \bibinfo
  {author} {\bibfnamefont {K.}~\bibnamefont {Nomura}},\ }\bibfield  {title}
  {\bibinfo {title} {Disorder effects in the anomalous hall effect induced by
  berry curvature},\ }\href {https://doi.org/10.1103/PhysRevB.72.045346}
  {\bibfield  {journal} {\bibinfo  {journal} {Phys. Rev. B}\ }\textbf {\bibinfo
  {volume} {72}},\ \bibinfo {pages} {045346} (\bibinfo {year}
  {2005})}\BibitemShut {NoStop}%
\bibitem [{\citenamefont {Yao}\ \emph {et~al.}(2004)\citenamefont {Yao},
  \citenamefont {Kleinman}, \citenamefont {MacDonald}, \citenamefont {Sinova},
  \citenamefont {Jungwirth}, \citenamefont {Wang}, \citenamefont {Wang},\ and\
  \citenamefont {Niu}}]{Yao2004}%
  \BibitemOpen
  \bibfield  {author} {\bibinfo {author} {\bibfnamefont {Y.}~\bibnamefont
  {Yao}}, \bibinfo {author} {\bibfnamefont {L.}~\bibnamefont {Kleinman}},
  \bibinfo {author} {\bibfnamefont {A.~H.}\ \bibnamefont {MacDonald}}, \bibinfo
  {author} {\bibfnamefont {J.}~\bibnamefont {Sinova}}, \bibinfo {author}
  {\bibfnamefont {T.}~\bibnamefont {Jungwirth}}, \bibinfo {author}
  {\bibfnamefont {D.-s.}\ \bibnamefont {Wang}}, \bibinfo {author}
  {\bibfnamefont {E.}~\bibnamefont {Wang}},\ and\ \bibinfo {author}
  {\bibfnamefont {Q.}~\bibnamefont {Niu}},\ }\bibfield  {title} {\bibinfo
  {title} {First principles calculation of anomalous hall conductivity in
  ferromagnetic bcc fe},\ }\href
  {https://doi.org/10.1103/PhysRevLett.92.037204} {\bibfield  {journal}
  {\bibinfo  {journal} {Phys. Rev. Lett.}\ }\textbf {\bibinfo {volume} {92}},\
  \bibinfo {pages} {037204} (\bibinfo {year} {2004})}\BibitemShut {NoStop}%
\bibitem [{\citenamefont {Hyart}\ \emph {et~al.}(2009)\citenamefont {Hyart},
  \citenamefont {Alexeeva}, \citenamefont {Mattas},\ and\ \citenamefont
  {Alekseev}}]{Hyart2009}%
  \BibitemOpen
  \bibfield  {author} {\bibinfo {author} {\bibfnamefont {T.}~\bibnamefont
  {Hyart}}, \bibinfo {author} {\bibfnamefont {N.~V.}\ \bibnamefont {Alexeeva}},
  \bibinfo {author} {\bibfnamefont {J.}~\bibnamefont {Mattas}},\ and\ \bibinfo
  {author} {\bibfnamefont {K.~N.}\ \bibnamefont {Alekseev}},\ }\bibfield
  {title} {\bibinfo {title} {Terahertz bloch oscillator with a modulated
  bias},\ }\href {https://doi.org/10.1103/PhysRevLett.102.140405} {\bibfield
  {journal} {\bibinfo  {journal} {Phys. Rev. Lett.}\ }\textbf {\bibinfo
  {volume} {102}},\ \bibinfo {pages} {140405} (\bibinfo {year}
  {2009})}\BibitemShut {NoStop}%
\bibitem [{\citenamefont {Esaki}\ and\ \citenamefont {Tsu}(1970)}]{Esaki1970}%
  \BibitemOpen
  \bibfield  {author} {\bibinfo {author} {\bibfnamefont {L.}~\bibnamefont
  {Esaki}}\ and\ \bibinfo {author} {\bibfnamefont {R.}~\bibnamefont {Tsu}},\
  }\bibfield  {title} {\bibinfo {title} {Superlattice and negative differential
  conductivity in semiconductors},\ }\href
  {https://doi.org/10.1147/rd.141.0061} {\bibfield  {journal} {\bibinfo
  {journal} {IBM Journal of Research and Development}\ }\textbf {\bibinfo
  {volume} {14}},\ \bibinfo {pages} {61} (\bibinfo {year} {1970})}\BibitemShut
  {NoStop}%
\bibitem [{\citenamefont {Beltram}\ \emph {et~al.}(1990)\citenamefont
  {Beltram}, \citenamefont {Capasso}, \citenamefont {Sivco}, \citenamefont
  {Hutchinson}, \citenamefont {Chu},\ and\ \citenamefont {Cho}}]{Beltram1990}%
  \BibitemOpen
  \bibfield  {author} {\bibinfo {author} {\bibfnamefont {F.}~\bibnamefont
  {Beltram}}, \bibinfo {author} {\bibfnamefont {F.}~\bibnamefont {Capasso}},
  \bibinfo {author} {\bibfnamefont {D.~L.}\ \bibnamefont {Sivco}}, \bibinfo
  {author} {\bibfnamefont {A.~L.}\ \bibnamefont {Hutchinson}}, \bibinfo
  {author} {\bibfnamefont {S.-N.~G.}\ \bibnamefont {Chu}},\ and\ \bibinfo
  {author} {\bibfnamefont {A.~Y.}\ \bibnamefont {Cho}},\ }\bibfield  {title}
  {\bibinfo {title} {Scattering-controlled transmission resonances and negative
  differential conductance by field-induced localization in superlattices},\
  }\href {https://doi.org/10.1103/PhysRevLett.64.3167} {\bibfield  {journal}
  {\bibinfo  {journal} {Phys. Rev. Lett.}\ }\textbf {\bibinfo {volume} {64}},\
  \bibinfo {pages} {3167} (\bibinfo {year} {1990})}\BibitemShut {NoStop}%
\bibitem [{\citenamefont {Mojarro}\ \emph {et~al.}(2021)\citenamefont
  {Mojarro}, \citenamefont {Carrillo-Bastos},\ and\ \citenamefont
  {Maytorena}}]{Mojarro2021}%
  \BibitemOpen
  \bibfield  {author} {\bibinfo {author} {\bibfnamefont {M.~A.}\ \bibnamefont
  {Mojarro}}, \bibinfo {author} {\bibfnamefont {R.}~\bibnamefont
  {Carrillo-Bastos}},\ and\ \bibinfo {author} {\bibfnamefont {J.~A.}\
  \bibnamefont {Maytorena}},\ }\bibfield  {title} {\bibinfo {title} {Optical
  properties of massive anisotropic tilted dirac systems},\ }\href
  {https://doi.org/10.1103/PhysRevB.103.165415} {\bibfield  {journal} {\bibinfo
   {journal} {Phys. Rev. B}\ }\textbf {\bibinfo {volume} {103}},\ \bibinfo
  {pages} {165415} (\bibinfo {year} {2021})}\BibitemShut {NoStop}%
\bibitem [{\citenamefont {Mojarro}\ \emph {et~al.}(2022)\citenamefont
  {Mojarro}, \citenamefont {Carrillo-Bastos},\ and\ \citenamefont
  {Maytorena}}]{Mojarro2022}%
  \BibitemOpen
  \bibfield  {author} {\bibinfo {author} {\bibfnamefont {M.~A.}\ \bibnamefont
  {Mojarro}}, \bibinfo {author} {\bibfnamefont {R.}~\bibnamefont
  {Carrillo-Bastos}},\ and\ \bibinfo {author} {\bibfnamefont {J.~A.}\
  \bibnamefont {Maytorena}},\ }\bibfield  {title} {\bibinfo {title} {Hyperbolic
  plasmons in massive tilted two-dimensional dirac materials},\ }\href
  {https://doi.org/10.1103/PhysRevB.105.L201408} {\bibfield  {journal}
  {\bibinfo  {journal} {Phys. Rev. B}\ }\textbf {\bibinfo {volume} {105}},\
  \bibinfo {pages} {L201408} (\bibinfo {year} {2022})}\BibitemShut {NoStop}%
\bibitem [{\citenamefont {Khomeriki}\ and\ \citenamefont
  {Flach}(2016)}]{Khomeriki2016}%
  \BibitemOpen
  \bibfield  {author} {\bibinfo {author} {\bibfnamefont {R.}~\bibnamefont
  {Khomeriki}}\ and\ \bibinfo {author} {\bibfnamefont {S.}~\bibnamefont
  {Flach}},\ }\bibfield  {title} {\bibinfo {title} {Landau-zener bloch
  oscillations with perturbed flat bands},\ }\href
  {https://doi.org/10.1103/PhysRevLett.116.245301} {\bibfield  {journal}
  {\bibinfo  {journal} {Phys. Rev. Lett.}\ }\textbf {\bibinfo {volume} {116}},\
  \bibinfo {pages} {245301} (\bibinfo {year} {2016})}\BibitemShut {NoStop}%
\bibitem [{\citenamefont {Ramya~Parkavi}\ \emph {et~al.}(2021)\citenamefont
  {Ramya~Parkavi}, \citenamefont {Chandrasekar},\ and\ \citenamefont
  {Lakshmanan}}]{Parkavi2021}%
  \BibitemOpen
  \bibfield  {author} {\bibinfo {author} {\bibfnamefont {J.}~\bibnamefont
  {Ramya~Parkavi}}, \bibinfo {author} {\bibfnamefont {V.~K.}\ \bibnamefont
  {Chandrasekar}},\ and\ \bibinfo {author} {\bibfnamefont {M.}~\bibnamefont
  {Lakshmanan}},\ }\bibfield  {title} {\bibinfo {title} {Stable bloch
  oscillations and landau-zener tunneling in a non-hermitian
  $\mathcal{PT}$-symmetric flat-band lattice},\ }\href
  {https://doi.org/10.1103/PhysRevA.103.023721} {\bibfield  {journal} {\bibinfo
   {journal} {Phys. Rev. A}\ }\textbf {\bibinfo {volume} {103}},\ \bibinfo
  {pages} {023721} (\bibinfo {year} {2021})}\BibitemShut {NoStop}%
\bibitem [{\citenamefont {Bouchard}\ and\ \citenamefont
  {Luban}(1995)}]{Bouchard1995}%
  \BibitemOpen
  \bibfield  {author} {\bibinfo {author} {\bibfnamefont {A.~M.}\ \bibnamefont
  {Bouchard}}\ and\ \bibinfo {author} {\bibfnamefont {M.}~\bibnamefont
  {Luban}},\ }\bibfield  {title} {\bibinfo {title} {Bloch oscillations and
  other dynamical phenomena of electrons in semiconductor superlattices},\
  }\href {https://doi.org/10.1103/PhysRevB.52.5105} {\bibfield  {journal}
  {\bibinfo  {journal} {Phys. Rev. B}\ }\textbf {\bibinfo {volume} {52}},\
  \bibinfo {pages} {5105} (\bibinfo {year} {1995})}\BibitemShut {NoStop}%
\bibitem [{\citenamefont {Zener}\ and\ \citenamefont
  {Fowler}(1934)}]{Zener1934}%
  \BibitemOpen
  \bibfield  {author} {\bibinfo {author} {\bibfnamefont {C.}~\bibnamefont
  {Zener}}\ and\ \bibinfo {author} {\bibfnamefont {R.~H.}\ \bibnamefont
  {Fowler}},\ }\bibfield  {title} {\bibinfo {title} {A theory of the electrical
  breakdown of solid dielectrics},\ }\href
  {https://doi.org/10.1098/rspa.1934.0116} {\bibfield  {journal} {\bibinfo
  {journal} {Proceedings of the Royal Society of London. Series A, Containing
  Papers of a Mathematical and Physical Character}\ }\textbf {\bibinfo {volume}
  {145}},\ \bibinfo {pages} {523} (\bibinfo {year} {1934})}\BibitemShut
  {NoStop}%
\bibitem [{\citenamefont {Vandenberghe}\ \emph {et~al.}(2010)\citenamefont
  {Vandenberghe}, \citenamefont {Sorée}, \citenamefont {Magnus},\ and\
  \citenamefont {Groeseneken}}]{Vandenberghe2010}%
  \BibitemOpen
  \bibfield  {author} {\bibinfo {author} {\bibfnamefont {W.}~\bibnamefont
  {Vandenberghe}}, \bibinfo {author} {\bibfnamefont {B.}~\bibnamefont
  {Sorée}}, \bibinfo {author} {\bibfnamefont {W.}~\bibnamefont {Magnus}},\
  and\ \bibinfo {author} {\bibfnamefont {G.}~\bibnamefont {Groeseneken}},\
  }\bibfield  {title} {\bibinfo {title} {{Zener tunneling in semiconductors
  under nonuniform electric fields}},\ }\href
  {https://doi.org/10.1063/1.3311550} {\bibfield  {journal} {\bibinfo
  {journal} {Journal of Applied Physics}\ }\textbf {\bibinfo {volume} {107}},\
  \bibinfo {pages} {054520} (\bibinfo {year} {2010})}\BibitemShut {NoStop}%
\bibitem [{\citenamefont {Biesenthal}\ \emph {et~al.}(2019)\citenamefont
  {Biesenthal}, \citenamefont {Kremer}, \citenamefont {Heinrich},\ and\
  \citenamefont {Szameit}}]{Biesenthal2019}%
  \BibitemOpen
  \bibfield  {author} {\bibinfo {author} {\bibfnamefont {T.}~\bibnamefont
  {Biesenthal}}, \bibinfo {author} {\bibfnamefont {M.}~\bibnamefont {Kremer}},
  \bibinfo {author} {\bibfnamefont {M.}~\bibnamefont {Heinrich}},\ and\
  \bibinfo {author} {\bibfnamefont {A.}~\bibnamefont {Szameit}},\ }\bibfield
  {title} {\bibinfo {title} {Experimental realization of
  $\mathcal{P}\mathcal{T}$-symmetric flat bands},\ }\href
  {https://doi.org/10.1103/PhysRevLett.123.183601} {\bibfield  {journal}
  {\bibinfo  {journal} {Phys. Rev. Lett.}\ }\textbf {\bibinfo {volume} {123}},\
  \bibinfo {pages} {183601} (\bibinfo {year} {2019})}\BibitemShut {NoStop}%
\bibitem [{\citenamefont {Yoshida}\ \emph {et~al.}(2019)\citenamefont
  {Yoshida}, \citenamefont {Otaki}, \citenamefont {Otaki},\ and\ \citenamefont
  {Fukui}}]{Yoshida2019}%
  \BibitemOpen
  \bibfield  {author} {\bibinfo {author} {\bibfnamefont {A.}~\bibnamefont
  {Yoshida}}, \bibinfo {author} {\bibfnamefont {Y.}~\bibnamefont {Otaki}},
  \bibinfo {author} {\bibfnamefont {R.}~\bibnamefont {Otaki}},\ and\ \bibinfo
  {author} {\bibfnamefont {T.}~\bibnamefont {Fukui}},\ }\bibfield  {title}
  {\bibinfo {title} {Edge states, corner states, and flat bands in a
  two-dimensional $\mathcal{PT}$-symmetric system},\ }\href
  {https://doi.org/10.1103/PhysRevB.100.125125} {\bibfield  {journal} {\bibinfo
   {journal} {Phys. Rev. B}\ }\textbf {\bibinfo {volume} {100}},\ \bibinfo
  {pages} {125125} (\bibinfo {year} {2019})}\BibitemShut {NoStop}%
\bibitem [{\citenamefont {Ghorai}\ \emph {et~al.}(2025)\citenamefont {Ghorai},
  \citenamefont {Das}, \citenamefont {Varshney},\ and\ \citenamefont
  {Agarwal}}]{Ghorai_prl2025}%
  \BibitemOpen
  \bibfield  {author} {\bibinfo {author} {\bibfnamefont {K.}~\bibnamefont
  {Ghorai}}, \bibinfo {author} {\bibfnamefont {S.}~\bibnamefont {Das}},
  \bibinfo {author} {\bibfnamefont {H.}~\bibnamefont {Varshney}},\ and\
  \bibinfo {author} {\bibfnamefont {A.}~\bibnamefont {Agarwal}},\ }\bibfield
  {title} {\bibinfo {title} {Planar hall effect in quasi-two-dimensional
  materials},\ }\href {https://doi.org/10.1103/PhysRevLett.134.026301}
  {\bibfield  {journal} {\bibinfo  {journal} {Phys. Rev. Lett.}\ }\textbf
  {\bibinfo {volume} {134}},\ \bibinfo {pages} {026301} (\bibinfo {year}
  {2025})}\BibitemShut {NoStop}%
\bibitem [{\citenamefont {Mandal}\ \emph {et~al.}(2024)\citenamefont {Mandal},
  \citenamefont {Sarkar}, \citenamefont {Das},\ and\ \citenamefont
  {Agarwal}}]{Mandal_prb2024}%
  \BibitemOpen
  \bibfield  {author} {\bibinfo {author} {\bibfnamefont {D.}~\bibnamefont
  {Mandal}}, \bibinfo {author} {\bibfnamefont {S.}~\bibnamefont {Sarkar}},
  \bibinfo {author} {\bibfnamefont {K.}~\bibnamefont {Das}},\ and\ \bibinfo
  {author} {\bibfnamefont {A.}~\bibnamefont {Agarwal}},\ }\bibfield  {title}
  {\bibinfo {title} {Quantum geometry induced third-order nonlinear transport
  responses},\ }\href {https://doi.org/10.1103/PhysRevB.110.195131} {\bibfield
  {journal} {\bibinfo  {journal} {Phys. Rev. B}\ }\textbf {\bibinfo {volume}
  {110}},\ \bibinfo {pages} {195131} (\bibinfo {year} {2024})}\BibitemShut
  {NoStop}%
\bibitem [{\citenamefont {Sarkar}\ and\ \citenamefont
  {Agarwal}(2026)}]{Sarkar_PIE2026}%
  \BibitemOpen
  \bibfield  {author} {\bibinfo {author} {\bibfnamefont {S.}~\bibnamefont
  {Sarkar}}\ and\ \bibinfo {author} {\bibfnamefont {A.}~\bibnamefont
  {Agarwal}},\ }\bibfield  {title} {\bibinfo {title} {Third-order rectification
  in centrosymmetric metals},\ }\href
  {https://doi.org/10.1088/2516-1083/ae5386} {\bibfield  {journal} {\bibinfo
  {journal} {Progress in Energy}\ }\textbf {\bibinfo {volume} {8}},\ \bibinfo
  {pages} {025004} (\bibinfo {year} {2026})}\BibitemShut {NoStop}%
\bibitem [{\citenamefont {Sarkar}\ \emph {et~al.}(2025)\citenamefont {Sarkar},
  \citenamefont {Das}, \citenamefont {Mandal},\ and\ \citenamefont
  {Agarwal}}]{Sarkar_njp2025}%
  \BibitemOpen
  \bibfield  {author} {\bibinfo {author} {\bibfnamefont {S.}~\bibnamefont
  {Sarkar}}, \bibinfo {author} {\bibfnamefont {S.}~\bibnamefont {Das}},
  \bibinfo {author} {\bibfnamefont {D.}~\bibnamefont {Mandal}},\ and\ \bibinfo
  {author} {\bibfnamefont {A.}~\bibnamefont {Agarwal}},\ }\bibfield  {title}
  {\bibinfo {title} {Light-induced nonlinear resonant spin magnetization},\
  }\href {https://doi.org/10.1088/1367-2630/ae1868} {\bibfield  {journal}
  {\bibinfo  {journal} {New Journal of Physics}\ }\textbf {\bibinfo {volume}
  {27}},\ \bibinfo {pages} {113503} (\bibinfo {year} {2025})}\BibitemShut
  {NoStop}%
\bibitem [{\citenamefont {Adak}\ \emph {et~al.}(2024)\citenamefont {Adak},
  \citenamefont {Sinha}, \citenamefont {Agarwal},\ and\ \citenamefont
  {Deshmukh}}]{Adak_NRM2024}%
  \BibitemOpen
  \bibfield  {author} {\bibinfo {author} {\bibfnamefont {P.~C.}\ \bibnamefont
  {Adak}}, \bibinfo {author} {\bibfnamefont {S.}~\bibnamefont {Sinha}},
  \bibinfo {author} {\bibfnamefont {A.}~\bibnamefont {Agarwal}},\ and\ \bibinfo
  {author} {\bibfnamefont {M.~M.}\ \bibnamefont {Deshmukh}},\ }\bibfield
  {title} {\bibinfo {title} {Tunable moiré materials for probing berry physics
  and topology},\ }\href {https://doi.org/10.1038/s41578-024-00671-4}
  {\bibfield  {journal} {\bibinfo  {journal} {Nature Reviews Materials}\
  }\textbf {\bibinfo {volume} {9}},\ \bibinfo {pages} {481–498} (\bibinfo
  {year} {2024})}\BibitemShut {NoStop}%
\bibitem [{\citenamefont {Longhi}(2014)}]{Longhi2014}%
  \BibitemOpen
  \bibfield  {author} {\bibinfo {author} {\bibfnamefont {S.}~\bibnamefont
  {Longhi}},\ }\bibfield  {title} {\bibinfo {title} {$\mathcal {PT}$-symmetric
  optical superlattices},\ }\href
  {https://doi.org/10.1088/1751-8113/47/16/165302} {\bibfield  {journal}
  {\bibinfo  {journal} {Journal of Physics A: Mathematical and Theoretical}\
  }\textbf {\bibinfo {volume} {47}},\ \bibinfo {pages} {165302} (\bibinfo
  {year} {2014})}\BibitemShut {NoStop}%
\bibitem [{\citenamefont {Butz}(2005)}]{10.1007/3-540-30924-1_7}%
  \BibitemOpen
  \bibfield  {author} {\bibinfo {author} {\bibfnamefont {T.}~\bibnamefont
  {Butz}},\ }\bibfield  {title} {\bibinfo {title} {The electric field gradient
  produced by a gaussian charge density distribution},\ }in\ \href@noop {}
  {\emph {\bibinfo {booktitle} {HFI/NQI 2004}}},\ \bibinfo {editor} {edited by\
  \bibinfo {editor} {\bibfnamefont {K.}~\bibnamefont {Maier}}\ and\ \bibinfo
  {editor} {\bibfnamefont {R.}~\bibnamefont {Vianden}}}\ (\bibinfo  {publisher}
  {Springer Berlin Heidelberg},\ \bibinfo {address} {Berlin, Heidelberg},\
  \bibinfo {year} {2005})\ pp.\ \bibinfo {pages} {41--46}\BibitemShut {NoStop}%
\bibitem [{\citenamefont {Vanderbilt}(2018)}]{vanderbilt2018berry}%
  \BibitemOpen
  \bibfield  {author} {\bibinfo {author} {\bibfnamefont {D.}~\bibnamefont
  {Vanderbilt}},\ }\href {https://books.google.co.in/books?id=vRNwDwAAQBAJ}
  {\emph {\bibinfo {title} {Berry Phases in Electronic Structure Theory:
  Electric Polarization, Orbital Magnetization and Topological Insulators}}},\
  Titolo collana\ (\bibinfo  {publisher} {Cambridge University Press},\
  \bibinfo {year} {2018})\BibitemShut {NoStop}%
\bibitem [{\citenamefont {Sundaram}\ and\ \citenamefont
  {Niu}(1999)}]{PhysRevB.59.14915}%
  \BibitemOpen
  \bibfield  {author} {\bibinfo {author} {\bibfnamefont {G.}~\bibnamefont
  {Sundaram}}\ and\ \bibinfo {author} {\bibfnamefont {Q.}~\bibnamefont {Niu}},\
  }\bibfield  {title} {\bibinfo {title} {Wave-packet dynamics in slowly
  perturbed crystals: Gradient corrections and berry-phase effects},\ }\href
  {https://doi.org/10.1103/PhysRevB.59.14915} {\bibfield  {journal} {\bibinfo
  {journal} {Phys. Rev. B}\ }\textbf {\bibinfo {volume} {59}},\ \bibinfo
  {pages} {14915} (\bibinfo {year} {1999})}\BibitemShut {NoStop}%
\bibitem [{\citenamefont {Cheng}(2013)}]{cheng2013quantum}%
  \BibitemOpen
  \bibfield  {author} {\bibinfo {author} {\bibfnamefont {R.}~\bibnamefont
  {Cheng}},\ }\href@noop {} {\bibinfo {title} {Quantum geometric tensor
  (fubini-study metric) in simple quantum system: A pedagogical introduction}}
  (\bibinfo {year} {2013})\BibitemShut {NoStop}%
\bibitem [{\citenamefont {de~Jesús Espinosa-Champo}\ and\ \citenamefont
  {Naumis}(2023)}]{Espinosa-Champo_2024}%
  \BibitemOpen
  \bibfield  {author} {\bibinfo {author} {\bibfnamefont {A.}~\bibnamefont
  {de~Jesús Espinosa-Champo}}\ and\ \bibinfo {author} {\bibfnamefont {G.~G.}\
  \bibnamefont {Naumis}},\ }\bibfield  {title} {\bibinfo {title}
  {Fubini–study metric and topological properties of flat band electronic
  states: the case of an atomic chain with s−p orbitals},\ }\href
  {https://doi.org/10.1088/1361-648X/acfbd1} {\bibfield  {journal} {\bibinfo
  {journal} {Journal of Physics: Condensed Matter}\ }\textbf {\bibinfo {volume}
  {36}},\ \bibinfo {pages} {015502} (\bibinfo {year} {2023})}\BibitemShut
  {NoStop}%
\bibitem [{\citenamefont {Graf}\ and\ \citenamefont
  {Pi\'echon}(2021)}]{PhysRevB.104.085114}%
  \BibitemOpen
  \bibfield  {author} {\bibinfo {author} {\bibfnamefont {A.}~\bibnamefont
  {Graf}}\ and\ \bibinfo {author} {\bibfnamefont {F.}~\bibnamefont
  {Pi\'echon}},\ }\bibfield  {title} {\bibinfo {title} {Berry curvature and
  quantum metric in $n$-band systems: An eigenprojector approach},\ }\href
  {https://doi.org/10.1103/PhysRevB.104.085114} {\bibfield  {journal} {\bibinfo
   {journal} {Phys. Rev. B}\ }\textbf {\bibinfo {volume} {104}},\ \bibinfo
  {pages} {085114} (\bibinfo {year} {2021})}\BibitemShut {NoStop}%
\bibitem [{\citenamefont {Ma}\ \emph {et~al.}(2010)\citenamefont {Ma},
  \citenamefont {Chen}, \citenamefont {Fan},\ and\ \citenamefont
  {Liu}}]{PhysRevB.81.245129}%
  \BibitemOpen
  \bibfield  {author} {\bibinfo {author} {\bibfnamefont {Y.-Q.}\ \bibnamefont
  {Ma}}, \bibinfo {author} {\bibfnamefont {S.}~\bibnamefont {Chen}}, \bibinfo
  {author} {\bibfnamefont {H.}~\bibnamefont {Fan}},\ and\ \bibinfo {author}
  {\bibfnamefont {W.-M.}\ \bibnamefont {Liu}},\ }\bibfield  {title} {\bibinfo
  {title} {Abelian and non-abelian quantum geometric tensor},\ }\href
  {https://doi.org/10.1103/PhysRevB.81.245129} {\bibfield  {journal} {\bibinfo
  {journal} {Phys. Rev. B}\ }\textbf {\bibinfo {volume} {81}},\ \bibinfo
  {pages} {245129} (\bibinfo {year} {2010})}\BibitemShut {NoStop}%
\end{thebibliography}%
\end{document}